\documentclass[11pt, a4paper]{article}

\usepackage[height=8.85in,width=6.4in]{geometry}

\usepackage{graphicx,rotating}     
\usepackage[bookmarksopen,colorlinks=true,linkcolor=light_blue,
citecolor=light_pink,urlcolor=dark_red,linktoc=all]{hyperref}

\usepackage{amsmath}
\usepackage{mathtools}
\usepackage{amsfonts}
\usepackage{amssymb}
\usepackage{slashed}
\usepackage{braket}
\usepackage{bm}
\usepackage{cite}
\usepackage{comment}
\usepackage{xcolor}
\usepackage[utf8]{inputenc}
\usepackage[footnotesize]{caption}

\usepackage{fourier}

\newcommand{\vev}[1]{ \left\langle {#1} \right\rangle }
\newcommand{\dd}{\mathrm{d}}

\newcommand{\abs}[1]{\left\vert {#1} \right\vert}
\newcommand{\Tr}{\text{Tr\,}}

\newcommand{\tilj}{\tilde{\jmath}}
\newcommand{\tilm}{\tilde{m}}
\def\normord#1{\mathop{:}\nolimits #1 \mathop{:}\nolimits}
\newcommand{\K}{K}

\makeatletter
\@addtoreset{equation}{section}

\makeatother

\usepackage{multicol}
\definecolor{dark_red}{rgb}{0.7, 0., 0.}
\definecolor{light_pink}{rgb}{1,0.4,0.4}
\definecolor{light_blue}{rgb}{0.284602,0.317763,0.963947}
\definecolor{cred}{RGB}{180,50,40} 
\definecolor{darkgreen}{RGB}{0, 100, 0}
\definecolor{desy_blue}{HTML}{009EE2}
\definecolor{desy_orange}{HTML}{FD8800}
\definecolor{forestgreen}{HTML}{228B22}
\definecolor{ochre}{HTML}{CCAA2B}

\begin{document}

\hypersetup{pageanchor=false}
\begin{titlepage}

\begin{center}

\hfill DESY 18-223\\
\hfill KEK-TH-2090\\

\vskip 1.in

{\Huge \bfseries 
Chiral Anomaly and Schwinger Effect\\
in Non-Abelian Gauge Theories\\
}
\vskip .8in

{\Large Valerie Domcke$^\lozenge$, Yohei Ema$^{\lozenge,\blacklozenge}$, Kyohei Mukaida$^\lozenge$, Ryosuke Sato$^\lozenge$}

\vskip .3in
\begin{tabular}{ll}
$^\lozenge$&\!\!\!\! \emph{DESY, Notkestra{\ss}e 85, D-22607 Hamburg, Germany}\\
$^\blacklozenge$&\!\!\!\! \emph{KEK Theory Center, Tsukuba 305-0801, Japan}
\end{tabular}

\end{center}
\vskip .6in

\begin{abstract}
\noindent
We study the production of chiral fermions in a background of a strong non-abelian gauge field with a non-vanishing Chern-Pontryagin density.
We discuss both pair production analogous to the Schwinger effect as well as asymmetric production through the chiral anomaly, sourced by the  Chern-Pontryagin density.
In abelian gauge theories one may nicely understand these processes by considering that the fermion dispersion relation forms discrete Landau levels.
Here we extend this analysis to a non-abelian gauge theory, considering an intrinsically non-abelian \textit{isotropic} and \textit{homogeneous} SU(2) gauge field background with a non-vanishing Chern-Pontryagin density. 
We show that the asymmetric fermion production, together with a non-trivial \textit{vacuum contribution}, correctly reproduces the chiral anomaly.
This indicates that the usual vacuum subtraction scheme, imposing normal ordering, fails in this case.
As a concrete example of this gauge field background,
we consider chromo-natural inflation.
Applying our analysis to this particular model, we compute the backreaction of the generated fermions on the gauge field background. This backreaction
receives contributions both from the vacuum through a Coleman-Weinberg-type correction and from the fermion excitations through an induced current.
\end{abstract}

\end{titlepage}

\tableofcontents
\thispagestyle{empty}
\renewcommand{\thepage}{\arabic{page}}
\renewcommand{\thefootnote}{$\natural$\arabic{footnote}}
\setcounter{footnote}{0}
\newpage
\hypersetup{pageanchor=true}

\section{Introduction}
\label{sec:intro}

Symmetries have been an invaluable guiding principle in the construction of the Standard Model (SM) of particle physics. On the one hand, gauge symmetries dictate the particle content and interactions. On the other hand, (approximate) global symmetries explain the lightness of scalars (such as the pion) as well as the (approximate) conservation of global charges (such as baryon $B$ and lepton $L$ number). In this context, a special role is played by global symmetries which are unbroken in the classical field theory, but violated by quantum corrections. Such `anomalously' broken symmetries explain \textit{e.g.}, the unexpectedly large pion decay rate~\cite{Adler:1969gk,Bell:1969ts}, the violation of $B+L$ through electroweak sphaleron processes~\cite{Klinkhamer:1984di}, and the chiral magnetic effect~\cite{Vilenkin:1980fu,Alekseev:1998ds,Fukushima:2008xe,Son:2012wh,Son:2012bg,Zyuzin:2012tv}. The anomaly equation states that the non-conservation of the corresponding chiral current (\textit{i.e.}, the difference of right-handed fermion current $J^\mu_\text{R}$ and left-handed fermion current $J^\mu_\text{L}$), is determined by the gauge field configuration entering the Chern-Pontryagin density $F \tilde F$~\cite{Fujikawa:1979ay,Fujikawa:1980eg},
\begin{equation}
 	\partial_\mu \left( J^\mu_\text{R} - J^\mu_\text{L} \right) =
	- \frac{1}{16 \pi^2}\, F_{\mu\nu} \tilde{F}^{\mu\nu}\,,
	\label{eq:anomaly_intro}
\end{equation}
where $F_{\mu \nu}$ ($\tilde F^{ \mu \nu})$ denotes the (dual) field strength tensor of the gauge field. In other words, certain gauge field configurations lead to an asymmetric fermion production. The result~\eqref{eq:anomaly_intro} can be elegantly proven in the path-integral formalism~\cite{Fujikawa:1979ay,Fujikawa:1980eg} and has been demonstrated to be exact to all orders in perturbation theory~\cite{Adler:1969er}.

In this paper we provide a microphysical derivation of the anomaly equation~\eqref{eq:anomaly_intro} in an SU$(2)$ gauge theory, based on solving the fermion equation of motion in a gauge field background. This task was performed for an abelian gauge theory in Ref.~\cite{Nielsen:1983rb}. There, the Lorentz force confined the fermion motion onto cylindrical orbits, leading a dispersion relation characterized by discrete Landau levels. The lowest of these Landau levels smoothly connects negative and positive energy states, and was identified as the source of the asymmetric fermion production accounted for by the anomaly equation. This microphysical understanding then allows for further results beyond reproducing the anomaly equation: the higher Landau levels, which due to their symmetry {under a parity transformation} do not contribute to the anomaly equation, instead allow for pair production of fermions and anti-fermions, analogous to Schwinger pair production in a strong electric field~\cite{Heisenberg:1935qt,Schwinger:1951nm}. In Ref.~\cite{Domcke:2018eki} these results were extended to account for the strong backreaction of the produced fermions on the gauge field background, with implications for axion inflation models and leptogenesis.

Extending this analysis to the non-abelian case, we encounter a number of significant differences. Firstly, non-abelian gauge fields allow for a non-vanishing, isotropic and homogenous gauge field background which sources a non-vanishing Chern-Pontryagin density. In this isotropic background, the fermion dispersion relation does not form discrete Landau levels, but instead we find an energy spectrum which is fully asymmetric between left- and right-handed fermions. Corresponding to the four degrees of freedom of a Dirac fermion SU$(2)$ doublet, we find one mode which smoothly connects negative and positive energy eigenstates as well as three gapped modes. Secondly, asymmetric fermion production in the gapless mode alone does not reproduce Eq.~\eqref{eq:anomaly_intro}. Instead, due to the asymmetric structure of the energy levels, there is an additional vacuum contribution, known to some as the eta invariant~\cite{Atiyah:1963zz,Atiyah:1968mp}. Thirdly, the pair production of fermions is a non-adiabatic process, arising from the time-dependent mixing of two of the energy eigenstates. And finally, compared to the abelian case, the backreaction of the induced fermion current on the background gauge field is less significant, at least for the exemplary non-adiabatic evolution considered here.

As a concrete example of these results, we turn to chromo-natural inflation (CNI)~\cite{Adshead:2012kp}. Here a pseudoscalar $\phi$ with a coupling to $F \tilde F$ is responsible for driving cosmic inflation, resulting in a strong homogeneous and isotropic gauge field background present during inflation. On the theoretical side, this model is attractive because it explains the flatness of the scalar potential by means of an approximate shift symmetry (respected by the coupling $\phi F \tilde F$) and because such shift-symmetric pseudoscalars (\textit{i.e.}, axion-like particles) arise numerously in string theory. On the phenomenological side, the prospect of strong gravitational wave production has received a lot of attention~\cite{Dimastrogiovanni:2012st,Dimastrogiovanni:2012ew,Adshead:2013qp,Adshead:2013nka}. Here, we compute the production of fermions charged under the SU$(2)$ gauge group employed in chromo-natural inflation in the presence of a chiral anomaly. In particular, we compute the backreaction on the gauge field background, which receives contributions from the vacuum (both adiabatic and non-adiabatic) and from the fermion excitations (both symmetric and asymmetric). 
{Similar studies have been carried out for scalar fields charged under the SU$(2)$ gauge group, finding that their backreaction is suppressed because all the modes become gapped~\cite{Lozanov:2018kpk}. 
A striking difference in our case is that the production of fermions through the chiral anomaly} {does not show this suppression, \textit{i.e.}, the operator equation~\eqref{eq:anomaly_intro} indicates the presence of at least one gapless mode.}
Nevertheless, we find these backreaction effects to be small, indicating that the non-linear effects of the non-abelian gauge field background are the dominant effect in constraining the gauge field growth. The fermion production during chromo-natural inflation may however be relevant for a subsequent phase of (p)reheating and/or baryogenesis.

The remainder of the paper is organized as follows. In Sec.~\ref{sec:pre} we specify our setup, deriving the equations of motion for the fermions, which set the foundation for the following computations. We also briefly introduce chromo-natural inflation as a concrete example featuring a strong gauge field background. The derivation of the anomaly equation is at the core of Sec.~\ref{sec:anomaly}, which includes the computation of vacuum contribution and of the contribution from the gapless fermion mode. In Sec.~\ref{sec:app_cni} we include non-adiabatic contributions as well as the backreaction on the gauge field background, focussing on the example of chromo-natural inflation. We conclude in Sec.~\ref{sec:conclusion}. Technical details are relegated to our four appendices. In particular, App.~\ref{sec:nandc} specifies our conventions and provides details on $CP$ transformation. App.~\ref{app:regularization} provides supplementary material on the derivation of the anomaly equation, in particular on the regularization of the vacuum contribution. App.~\ref{sec:general} extends the discussion of the main text (which focuses on fermions in the fundamental representation) to general fermion representations. Finally, App.~\ref{sec:bogogo} contains an analytical derivation of the Bogolyubov coefficients responsible for fermion pair production.

\section{Fermions in a non-abelian gauge field background}
\label{sec:pre}

\subsection{Setup and motivation}
\label{sec:model}

\paragraph{Model.}
As a minimal setup for fermion production in non-abelian gauge theories in the presence of a chiral anomaly, we consider massless chiral fermions charged under an SU$(2)$ gauge theory.
To see the difference between the left (L) - and right (R) -handed fermions explicitly, we include both of them, working in $4$-spinor notation, $\psi = \psi_\text{L} + \psi_\text{R}$.
Our starting point is the following action:
\begin{align}
	S_\psi = \int \dd^4 x\, \overline{\psi} \left( i \slashed{\partial} + \slashed{A}^a T^a  \right) \psi \,,
	\label{eq:S_psi}
\end{align}
where $T^a$ is a generator of the SU$(2)$ gauge group and $A^a_\mu$ represents an SU$(2)$ gauge field.\footnote{Throughout this paper, $a,b,.. = \{1,2,3\}$ denote gauge indices, $\mu, \nu,.. = \{0..3\}$ denote Lorentz indices and $i,j,.. = \{1..3\}$ denote spatial indices.  } 
Throughout this paper, we denote $x^0 = \eta$.  {Since the fermion is conformally coupled in this theory, our analysis applies both to flat space, characterized by the Minkowski metric, $(\eta_{\mu \nu}) = \text{Diag}(1,-1,-1,-1)$, as well as to an expanding universe described by the FLRW metric, $(g_{\mu \nu}) = \text{Diag}(a^2,-a^2,-a^2,-a^2)$ with $a(\eta)$ denoting the cosmic scale factor and $\eta$ referring to conformal time. See Sec.~\ref{sec:cni} for details.}

Classically, the left- and right-handed fermion currents are separately conserved.
However, once quantum processes are included, the axial combination is broken, which is known as the chiral anomaly: 
\begin{align}
	\partial_\mu \left( J^\mu_\text{R} + J^\mu_\text{L} \right) &= 0\,, \label{eq:vector_current}\\[.5em]
	\partial_\mu \left( J^\mu_\text{R} - J^\mu_\text{L} \right) &=
	- \frac{1}{8 \pi^2}\, T(\textbf{r})\,  F^a_{\mu\nu} \tilde{F}^{a \; \mu\nu}\,, \label{eq:axial_current}
\end{align}
where the left/right-handed currents are defined by
\begin{align}
	J^\mu_H \equiv \overline \psi \gamma^\mu \mathcal{P}_H \psi \,.
\end{align}
Here the subscript $H = \{\text{L}, \text{R} \}$ indicates helicity, with the corresponding projection operator  defined by $\mathcal{P}_\text{R/L} \equiv (1 \pm \gamma_5)/2$. {$F_{\mu \nu}^a$ denotes the field strength tensor of the SU$(2)$ gauge group, $F_{\mu \nu}^a = \partial_\mu A_\nu^a - \partial_\nu A_\mu^a +  \epsilon^{abc} A_\mu^b A_\nu^c$}, with the dual field strength defined as $\tilde F^{a \; \mu\nu} \equiv \epsilon^{\mu\nu\rho\sigma} F^a_{\rho\sigma}/2$. The convention of the total antisymmetric tensor is fixed by $\epsilon^{0123} = +1$.
Note that $2 T(\textbf{r})$ is the Dynkin index of a representation ($\textbf{r}$) defined by $\Tr (T^a T^b) = T (\textbf{r})\, \delta^{ab}$. For a fundamental representation ($\textbf{2}$) we have $T(\textbf{2}) = 1/2$.
In the main text we will focus on fermions in the fundamental representation. See appendix~\ref{sec:general} for a general representation.

For later convenience we define charges associated to these currents,
\begin{align}
	Q_H (\eta) \equiv \int \dd^3 x\, J^0_H (\eta, \bm{x})\,.
\end{align}
The conservation laws in terms of these charges are\footnote{
	One may drop $\partial_i \langle J_H^i \rangle$ because we assume the translational invariance throughout this paper, which implies $\langle J_H^i (\eta, \bm{x}) \rangle = \langle J_H^i (\eta, \bm{0}) \rangle$.
}
\begin{align}
	\frac{\partial}{\partial \eta} \left( \vev{Q_\text{R}} + \vev{Q_\text{L}}  \right) &= 0, \label{eq:vector}\\
	\frac{\partial}{\partial \eta} \left( \vev{Q_\text{R}} - \vev{Q_\text{L}}  \right) &= 
	- \frac{1}{8 \pi^2}\, T(\textbf{r})\, \int \dd^3 x\,  \vev{F^a_{\mu\nu} \tilde{F}^{a \mu\nu}}\,,
	\label{eq:axial}
\end{align}
where $\vev{ \bullet }$ denotes an expectation value of a given state.

\paragraph{Motivation.}
Generally speaking, fermions charged under a gauge group can be generated in the presence of a strong gauge field.
One may study this process by solving the equation of motion for fermions in the background of this gauge field.
Before turning to fermion production in an intrinsically non-abelian background, let us first re-call the situation in a (quasi-)Abelian background.
A classic example is studied in Ref.~\cite{Nielsen:1983rb}: Fermion production in parallel electric and magnetic fields in abelian gauge field theory.
There the dispersion relations of fermions form discrete Landau levels since the magnetic field restricts the motion of fermions transverse to the magnetic field by the Lorentz force.
Taking the electromagnetic fields parallel to the $z$-axis, the lowest Landau level, which represents the mode (anti)parallel to the magnetic field, has the following dispersion relation: $\omega = \pm p_z$ for the right- and left-handed fermions, respectively.
Thus the positive and negative energy states are smoothly connected.
Since fermions get accelerated by the electric field, $\dot p_z = gQE$, the highest lying negative energy states are converted into positive energy states for the right-handed fermions (and vice versa for the left-handed fermions), leading to a chiral asymmetry.
Ref.~\cite{Nielsen:1983rb} shows explicitly that the resulting chiral asymmetry computed in this way, \textit{i.e.}, by solving the equation of motion, is consistent with the anomaly equation.

Our primary motivation is to extend this analysis to a non-abelian gauge field. 
One may \textit{e.g.}, consider the following background of the non-abelian gauge field~\cite{Tanji:2011di}:\footnote{
	Note that this configuration gives \textit{homogeneous} electric and magnetic fields pointing the $z$ direction. As a result, they do not break the translational invariance and hence we may use Eqs.~\eqref{eq:vector} and \eqref{eq:axial}.
}
\begin{align}
	A^a_\mu = \overline{A}_\mu n^a,
	\quad \overline{A}_\mu = \left(0, 0, -Bx,  E \eta \right),
\end{align}
where $n^a$ is an arbitrary constant unit vector.
This leads to {homogeneous} color electric/magnetic fields pointing along the $z$-axis.
This configuration is, however, essentially the same as abelian gauge field case.
The unit vector, $n^a$, projects the non-abelian gauge group onto its U$(1)$ subgroup and hence one may apply the analysis of the abelian gauge field straightforwardly.
One may take $(n^a) = (0,0,1)$ without loss of generality.
The effective charges of fermions with respect to the \textit{abelian} gauge field, $\overline{A}_\mu$, are given by $m = - j, - j + 1, \dots, j-1, j$ for a $\textbf{2j+1}$ representation of SU$(2)$.
{Now it is clear that the computation is exactly the same as the abelian case. All one has to do is to sum over all the fermions, namely a summation over the charge squared, which is nothing but the Dynkin index:}
\begin{align}
	\sum_{m = -j}^j m^2 = \frac{1}{3} (2 j +1)(j+1)j  = T (\textbf{2j+1}) \,.
\end{align}
Therefore we reproduce the anomaly equation \eqref{eq:axial}.

In this paper we consider a more intriguing configuration which cannot be achieved in abelian gauge theory, \textit{i.e.}, an intrinsically non-abelian gauge field configuration.
{Contrary to an abelian gauge theory, an SU$(2)$ gauge theory allows for a non-vanishing homogeneous and isotropic gauge field background. Up to spatial rotations and gauge transformations, this is uniquely given by  (see \textit{e.g.},~\cite{Verbin:1989sg,Maleknejad:2012fw,Domcke:2018rvv})}
\begin{align}
	A^a_0 = 0\,, \quad
	A^a_i =  -f (\eta)\, \delta^a_i\,,
	\label{eq:gauge_bkg_hom}
\end{align}
{where we have imposed temporal gauge.}  
Interestingly, this configuration gives rise to a non-vanishing Chern-Pontryagin density,
\begin{align}
	E^{a i} = - \partial_\eta A^{ai} = - \partial_\eta f\, \delta^{ai}\, , \quad
	B^{ai} = \frac{1}{2} \epsilon^{ijk} F{{^a}_j}^k = - f^2 \delta^{ai}\,,
	\quad 
	F^a_{\mu\nu} \tilde{F}^{a\mu\nu}
	= - 4 E^{ai} B^{ai} 	
	= - 12 f' f^2\,.
	\label{eq:topo_dens}
\end{align}
{The non-vanishing color magnetic field arising from the homogeneous vector potential is a clear indicator of the intrinsically non-abelian nature of this phenomenon. Due to homogeneity and isotropy, we no longer expect the fermion dispersion relation to be described by Landau levels, hence we expect the implementation of the anomaly equation on the microphysical level to be qualitatively different than in the abelian case. }
{Note that the configuration~\eqref{eq:gauge_bkg_hom} can be realized for {any gauge} group which has an SU$(2)$ subgroup.}

In the remainder of this paper we study the fermion production by solving the equation of motion for a fermion in the non-trivial homogeneous and isotropic gauge field background~\eqref{eq:gauge_bkg_hom}.
A primary example of this gauge field configuration is chromo-natural inflation, see Sec.~\ref{sec:cni}.
In Sec.~\ref{sec:app_cni}, we apply our analysis to chromo-natural inflation and discuss the backreaction of generated fermions on the gauge field.

\subsection{Chromo-natural inflation \label{sec:CNI}}
\label{sec:cni}

Well-studied applications of the gauge field background~\eqref{eq:gauge_bkg_hom} arise \textit{e.g.}, in models of cosmic inflation employing SU(2) gauge fields: in `gauge-flation'~\cite{Maleknejad:2011jw,Maleknejad:2011sq}    (see also~\cite{Adshead:2012qe,SheikhJabbari:2012qf}) a non-trivial isotropic gauge field background was shown to support a phase of cosmic inflation, in `chromo-natural infation'~\cite{Adshead:2012kp} the presence of such a gauge field background was shown act as an effective friction term in the inflaton dynamics. In both cases, the gauge field fluctuations around this background, exponentially enhanced through a tachyonic instability, can source a sizable gravitational wave background~\cite{Dimastrogiovanni:2012st,Dimastrogiovanni:2012ew,Adshead:2013qp,Adshead:2013nka}.
\footnote{While the minimal model is by now disfavoured by CMB observations, consistency with the data can be achieved by employing different inflation potentials~\cite{Caldwell:2017chz,DallAgata:2018ybl}, by enlarging the field content of the model~\cite{Dimastrogiovanni:2016fuu,McDonough:2018xzh}, by considering a spontaneously broken gauge symmetry~\cite{Adshead:2016omu} or by taking into account the dynamical evolution of the gauge field background from Bunch Davies initial conditions~\cite{Domcke:2018rvv}.}

\paragraph{Action.}

In the following we focus on the example of chromo-natural inflation, considering
the following action,
\begin{align}
	S = S_\text{EH} + S_\psi + S_\text{CNI} \,, 
\end{align}
where $S_\text{EH}$ is the usual Einstein-Hilbert action, $S_\psi$ denotes the action for a massless fermion and $S_\text{CNI}$ describes a pseudoscalar $\phi$ coupled to the SU$(2)$ gauge fields. After identifying $\phi$ as the inflaton, the latter describes chromo-natural inflation (CNI), and is given by
\begin{align}
	S_\text{CNI} = \int \dd^4 x\, 
	\Bigg\{ \sqrt{-g}\left[
		\frac{g^{\mu\nu}}{2} \partial_\mu \phi \partial_\nu \phi - V (\phi)
		- \frac{1}{4 g^2} g^{\mu \rho} g^{\nu \sigma} \hat F^a_{\mu \nu} \hat F^a _{\rho \sigma}
	\right] 
	+ \frac{\phi}{16 \pi^2 f_a} \hat F^a_{\mu \nu} \hat{\tilde F}^{a\mu \nu}
	\Bigg\}\,,
\end{align}
where $V(\phi)$ is a potential and $g$ denotes the SU$(2)$ gauge coupling.\footnote{
The gauge coupling, $g$, should not be confused with the FLRW metric $g_{\mu\nu}$ or it's determinant $\sqrt{-g} = \sqrt{- \text{det} (g_{\mu \nu})} $.
} 
The coupling to the Chern-Pontryagin density $\hat F \hat{\tilde F}$ is determined by the axion decay constant $f_a$, which may be interpreted as the cut-off scale of the effective field theory. {Here, to distinguish from the formulation in co-moving coordinates in the previous section, we have used ``$ \hat \bullet $'' to denote quantities in physical coordinates.}

The gauge field is conformal and hence one may factor out the expansion of the Universe 
by the following field redefinition:
\begin{align}
	(A_\mu^a) \equiv (\hat A_0^a, - \bm{A}^a) = (\hat A_\mu^a)\,, \quad
	(A^{a \mu}) \equiv (\hat A_0^a, \bm{A}^a) = a^2 (\hat A^{a \mu})\,,
\end{align}
with $a$ denoting the cosmic scale factor. 
Note that we raise/lower indices of the rescaled field, $A^a_\mu$, by the flat metric, $(\eta_{\mu\nu}) = \text{Diag} \, (1,-1,-1,-1)$, while those of the original field, $\hat A^a_\mu$, are raised/lowered by the FLRW metric, $(g_{\mu\nu}) = \text{Diag}\, (a^2,-a^2,-a^2,-a^2)$.
In this conformal basis our action simplifies to
\begin{align}
	S_\text{CNI}
	=
	\int \dd^4 x \Bigg\{ \sqrt{-g}
	\left[
		\frac{g^{\mu\nu}}{2} \partial_\mu \phi \partial_\nu \phi - V (\phi)
	\right]
		- \frac{1}{4 g^2} F^a_{\mu \nu} F^{a \mu\nu}
	+ \frac{\phi}{16 \pi^2 f_a} F^a_{\mu \nu} \tilde F^{a \mu \nu}
	\Bigg\}\,.
\end{align}
The massless fermion is also conformally invariant, 
and hence the fermion action in the FLRW background can be recast as (see, \textit{e.g.}, \cite{Domcke:2018eki})
\begin{align}
	 S_\psi
		=
		\int \dd^4 x\;
		\overline \psi  \gamma^\mu \left( i {\partial}_\mu 	+ {A}^a_\mu T^a \right) \psi\,,
\end{align}
where the gamma matrices here are defined on the flat coordinates: 
$\{\gamma^\mu, \gamma^\nu \} = 2 (\eta^{\mu\nu})\,$.
Following the derivation of Ref.~\cite{Fujikawa:1979ay,Fujikawa:1980eg} (see also \cite{Peskin:1995ev}), 
it is clear that the presence of $\phi$ and in particular its coupling to the Chern-Pontryagin density does not modify the chiral anomaly equation. 
Thus the usual equations for the vector and axial current, respectively as in Eqs.~\eqref{eq:vector_current} and \eqref{eq:axial_current}, remain valid.

\paragraph{Background equation of motion.}
In the presence of Chern-Simons type coupling $\phi F \tilde F$ with $\dot \phi \neq 0$, non-abelian gauge groups support the nontrival homogeneous isotropic solution specified in  Eq.~\eqref{eq:gauge_bkg_hom},
\begin{align}
	A^a_0 = 0, \quad
	A^a_i =  -f (\eta)\, \delta^a_i = -a(\eta) \hat f(\eta)\, \delta^a_i\,.
	\label{eq:gauge_bkg}
\end{align}
{Neglecting a possible backreaction arising from the fermions in the theory (to which we will return in Sec.~\ref{sec:app_cni}),} the equation of motion for this homogeneous component in a quasi de-Sitter background, $a(\eta) = 1/(- H \eta)$, reads
\begin{align}
 	f''(\eta) + 2 f^3(\eta) - \frac{2 \xi}{(- \eta)} f^2(\eta)  = 0 \,,
	\qquad \text{with} \quad 
	 \xi \equiv \frac{\alpha \dot \phi}{2 \pi f_a H} \,.
 	\label{eq:eom_f}
\end{align}
Here w.l.o.g.\ we have assumed $\dot \phi > 0$. We note that under a parity transformation, $\dot \phi \mapsto - \dot \phi$ and $f \mapsto - f$. 
Taking $\xi$ to be a constant parameter, {consistent with the slow-roll approximation $|\ddot \phi| \ll |3 H \dot \phi|, |V'(\phi)|$,}
Eq.~\eqref{eq:eom_f} admits three asymptotic solutions, $g f(\eta) = c_i \xi/(- \eta)$ with~\cite{Domcke:2018rvv} 
\begin{align}
 c_0 = 0 \,, \quad c_1 = \frac{1}{2}\left(1 - \sqrt{1 - 4 / \xi^2} \right) \,, \quad c_2 = \frac{1}{2}\left(1 + \sqrt{1 - 4 / \xi^2} \right) \,.
 \label{eq:solutions}
\end{align}
The $c_0$- and $c_2$-solutions are attractor solutions, and are obtained as the asymptotic state of general oscillatory solutions. The $c_1$-solution represents a local maximum separating the $c_0$ and the $c_2$ regime. 
Note that the $c_1$- and $c_2$-solutions are only possible for $\xi >  2$. Intuitively, one may think of $f(\eta)$ as a scalar degree of freedom in an effective time-dependent potential, which for $\xi > 2$ features a local minimum associated with $c_0$ and a global minimum associated with $c_2$. 

Of particular interest to us is the $c_2$-solution, which represents a non-trivial attractor solution for homogeneous isotropic SU$(2)$ gauge fields in quasi de-Sitter space.
When specifying the explicit form of $f(\eta)$ in Sec.~\ref{sec:app_cni}, we will choose this global minimum of CNI, \textit{i.e.},
\begin{align}
 f(\eta) = \frac{c_2 \xi}{- \eta} \simeq \frac{\xi}{- \eta} \,, \quad \hat f(\eta) = c_2 \xi H \simeq  \xi H \,,
 \label{eq:gauge_background}
\end{align}
where the last equality holds for $\xi \gg 2$. We note that in quasi de-Sitter space, $H \simeq$~const.\ and $\dot \phi \simeq$~const., the quantity $\hat f$ is approximately constant. {Expanding around the gauge field background~\eqref{eq:gauge_background}, one mode of the gauge field fluctuations acquires a tachyonic mass, leading to a strong non-perturbative production of this gauge field mode~\cite{Dimastrogiovanni:2012st,Dimastrogiovanni:2012ew,Adshead:2013qp,Adshead:2013nka,Domcke:2018rvv}. Since the main focus of this paper is fermion production, we will mostly ignore this instability in the following. We note that at least in some parts of the parameter space, the homogeneous gauge field background safely dominates over this non-perturbative contribution~\cite{Adshead:2016omu}.}

In the following we will study charged fermion production in the gauge field background~\eqref{eq:gauge_bkg_hom}. In Sec.~\ref{sec:app_cni}, where we need to specify the explicit time evolution of this gauge field background in order to study non-adiabatic processes, we will resort to chromo-natural inflation, Eq.~\eqref{eq:gauge_background}, as a prime example.

\subsection{Basic ingredients}
\label{sec:basic}
\paragraph{Equation of motion.}
To study the fermion production, we will solve the fermion equation of motion in this gauge field background. After a Fourier transformation,
\begin{align}
	\psi_\text{L/R} (\eta, \bm{x}) = 
	\int \frac{\dd^3 k}{(2 \pi)^{3/2}}\, e^{i \bm{k} \cdot \bm{x}} \psi_\text{L/R} (\eta, \bm{k}) \,,
\end{align}
this equation of motion is given by
\begin{align}
	0 = \left[ i \partial_\eta \pm \bm{\sigma} \cdot \bm{k} \pm f(\eta)\, \bm{\sigma} \cdot \bm{T} \right] \psi_\text{L/R} (\eta, \bm{k})\,,
	\label{eq:eom_fermion0}
\end{align}
with the Pauli matrices $\bm{\sigma}$ acting on the spin indices while the SU$(2)$ generators $T^a$ acting on the gauge indices of $\psi$. 
{Thanks to rotational invariance, we can take $\bm{k}$ along the z-direction without loss of generality. The eigenbases of the spin ($\chi^{(\pm)}_{\bm k}$) and gauge ($t^{(\pm)}_{\bm k}$) degrees of freedom then obey}
\begin{align}
	\left( \hat{\bm{k}} \cdot \bm{\sigma} \right) \chi^{(\pm)}_{\bm k} = \pm \chi^{(\pm)}_{\bm k},\quad
	\left( \hat{\bm{k}} \cdot \bm{T} \right) t^{(\pm)}_{\bm k} = \pm \frac{1}{2} t^{(\pm)}_{\bm k} \,,
\end{align}
with $\hat{\bm k} = \hat e_z$ and
\begin{align}
 {\chi^{(+)}_{\bm k}, t^{(+)}_{\bm k} = \begin{pmatrix} 1 \\ 0 \end{pmatrix} \,, \quad 
  \chi^{(-)}_{\bm k}, t^{(-)}_{\bm k} = \begin{pmatrix} 0 \\ 1 \end{pmatrix} \,. }
  \label{eq:basis_z}
\end{align}
One may expand the wave function of the fermion in terms of a product of these polarization vectors,\footnote{{Compared to the discussion for general representations in App.~\ref{sec:general}, we use $m = \pm 1$ instead of $m = \pm 1/2$ for the discussion of the fundamental representation in the main text to ease the notation.}}
\begin{align}
	\psi_\text{L/R} (\eta, \bm{k}) = \sum_{s,m=\pm} \psi_\text{L/R}^{(s,m)} (\eta, \bm{k}) \, \chi_{\bm k}^{(s)} t_{\bm k}^{(m)} \,.
	\label{eq:expnd_pol}
\end{align}
In this basis the equation of motion~\eqref{eq:eom_fermion0} is simplified significantly.
Noticing the following relation, $\bm{\sigma} \cdot \bm{T} = (\sigma_+ T_- + \sigma_- T_+)/2
+ (\hat{\bm{k}} \cdot \bm{\sigma}) (\hat{\bm{k}} \cdot \bm{T})$
with $\sigma_\pm$ and $T_\pm$ denoting the respective ladder operators,\footnote{$\sigma_\pm \chi_k^{(\mp)} = \chi_k^{(\pm)}$, $\sigma_\pm \chi_k^{(\pm)} = 0$ and $T_\pm$ analogously.}
one may easily see that the lowest mode, $\psi_\text{L/R}^{(-,-)}$, and highest mode, $\psi_\text{L/R}^{(+,+)}$, are decoupled,
while the other modes, $\psi_\text{L/R}^{(+,-)}$ and $\psi_\text{L/R}^{(-,+)}$ get mixed:
\begin{align}
	0 &= \left[ i \partial_\eta \pm \left(k + \frac{f(\eta)}{2} \right) \right] \psi_\text{L/R}^{(+,+)} (\eta, \bm{k})\,, \label{eq:++} \\[.5em]
	0 &= \left[ i \partial_\eta \mp \left(k - \frac{f(\eta)}{2} \right) \right] \psi_\text{L/R}^{(-,-)} (\eta, \bm{k})\,, \label{eq:--} \\[.5em]
	0 &= 
	\left[
	i \partial_\eta
	 \pm k \left(\begin{array}{cc}
	1 &    \\
	 & -1 \end{array}\right) 
	\pm \frac{f(\eta)}{2} \left(\begin{array}{cc}
	-1 & 2   \\
	 2 & -1 \end{array}\right) 
	\right]
	\left(\begin{array}{c}
	\psi_\text{L/R}^{(+,-)} (\eta, \bm{k}) \\
	\psi_\text{L/R}^{(-,+)} (\eta, \bm{k}) \end{array}\right)\,,
	\label{eq:mixed}
\end{align}
where $k \equiv \vert \bm{k}\vert$.
This mixing structure is easily understood once we note that 
the diagonal part of the SU(2) gauge and SO(3) spatial rotational symmetries, and hence $s+m$, 
is conserved in our gauge field configuration.

\paragraph{Energy eigenbasis for constant gauge field background.}
Before discussing the time evolution of $f (\eta)$,
let us discuss a constant gauge field background $f$ and see how this modifies the dispersion relation.
For a constant $f$, one may easily solve Eqs.~\eqref{eq:++}, \eqref{eq:--}, and \eqref{eq:mixed}. The solutions to these first order differential equations will be of the type $\psi \propto \exp(- i \omega(k) \eta)$, where we associate positive frequencies $\omega(k) > 0$ with particles and negative frequencies $\omega(k) < 0$ with anti-particles. More precisely, we find the dispersion relations:
\begin{align}
	\omega_\text{L/R}^{(+1)} &= \mp \left( k + \frac{f}{2} \right)\,, \label{eq:disp_++} \\[.5em]
	\omega_\text{L/R}^{(-1)} &= \pm \left( k - \frac{f}{2} \right)\,, \label{eq:disp_--} \\[.5em]
	\omega_\text{L/R}^{(0;1)} & = \mp \left( \sqrt{k^2 + f^2} - \frac{f}{2} \right)\,, \label{eq:disp_01} \\[.5em]	
	\omega_\text{L/R}^{(0;2)} & = \pm \left( \sqrt{k^2 + f^2} + \frac{f}{2} \right)\,. \label{eq:disp_02}
\end{align}
corresponding to four states at any given value of $k$ for each chirality, as expected for a fermion doublet. 
{Here the (first argument of the) superscript indicates the eigenvalue of the total spin in the $z$-direction, 
\textit{i.e.}, $(s+m)/2$.}
To obtain Eq.~\eqref{eq:disp_01} and \eqref{eq:disp_02}, 
we have diagonalized Eq.~\eqref{eq:mixed} by the rotation matrix $O$ which is defined by
\begin{align}
	O (\kappa) = 
	\left(\begin{array}{cc}
	\frac{1}{\sqrt{2}} \frac{\kappa}{\sqrt{\kappa^2}} \sqrt{1 + \frac{1}{\sqrt{1 + 1 / \kappa^2}}} & \frac{1}{\sqrt{2}}\sqrt{1 - \frac{1}{\sqrt{1 + 1 / \kappa^2}}} \\
	-\frac{1}{\sqrt{2}}\sqrt{1 - \frac{1}{\sqrt{1 + 1 / \kappa^2}}} & \frac{1}{\sqrt{2}} \frac{\kappa}{\sqrt{\kappa^2}} \sqrt{1 + \frac{1}{\sqrt{1 + 1 / \kappa^2}}} \end{array}\right)\,,
	\label{eq:rot}
\end{align}
so that
\begin{align}
	\left(\begin{array}{c}
	\psi_\text{L/R}^{(0;1)} (\eta, \bm{k}) \\
	\psi_\text{L/R}^{(0;2)} (\eta, \bm{k}) \end{array}\right) = 
	O(k/f)\,
	\left(\begin{array}{c}
	\psi_\text{L/R}^{(+,-)} (\eta, \bm{k}) \\
	\psi_\text{L/R}^{(-,+)} (\eta, \bm{k}) \end{array}\right)\,.
\end{align}
\begin{figure}
\centering
\includegraphics[width=0.49 \textwidth]{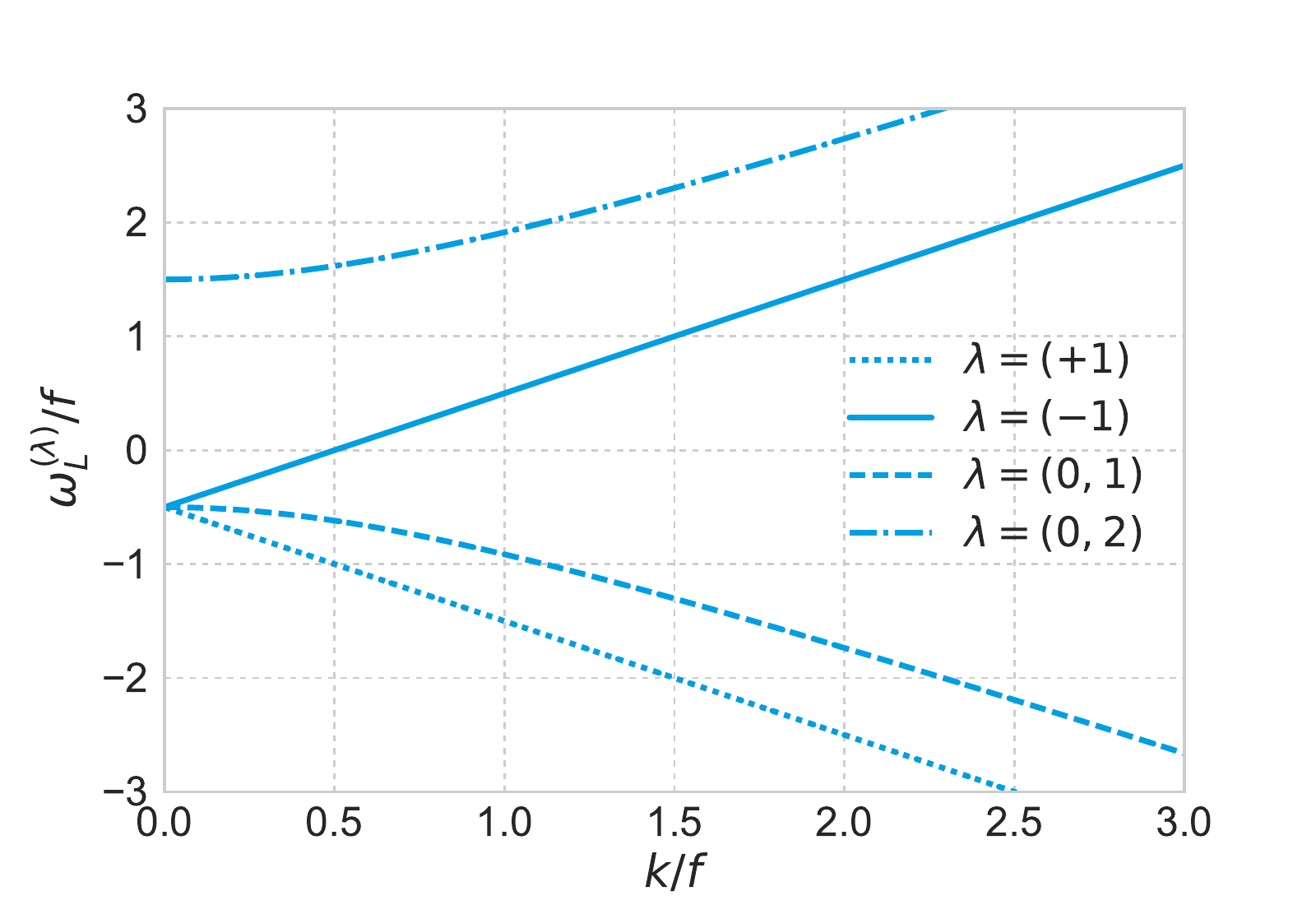}
\includegraphics[width=0.49 \textwidth]{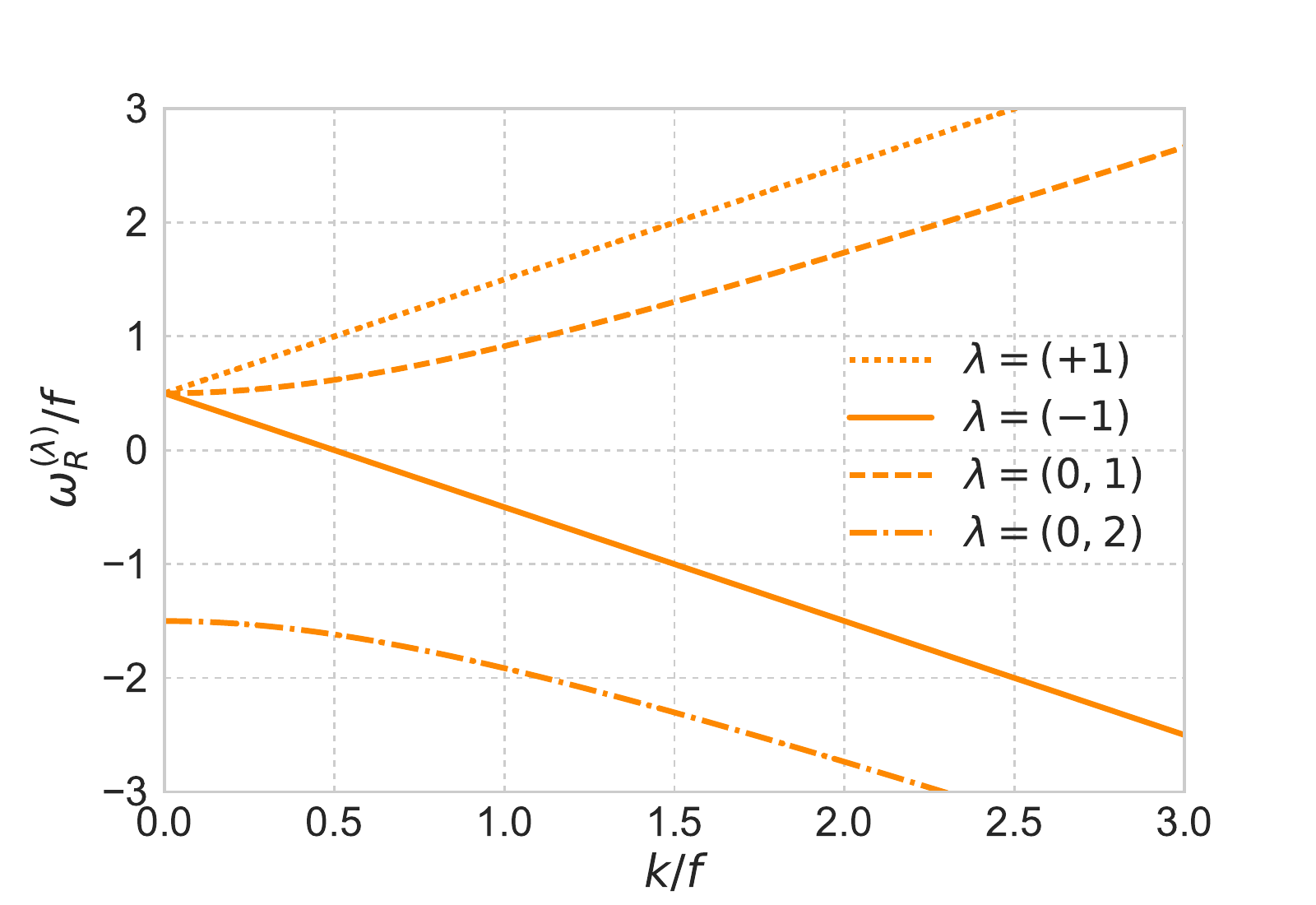}
\caption{
	\textbf{Left panel}: Dispersion relation for the left-handed fermion.
	One mode, $\lambda = (-1)$, crosses the zero energy while the other three modes are gapped.
	\textbf{Right panel}: Dispersion relation for the right-handed fermion.  It is obtained from flipping the sign of the energy for the left-handed fermion. 
}
\label{fig:dispersion}
\end{figure}

The dispersion relation is also shown in Fig.~\ref{fig:dispersion}.
There is one mode which smoothly connects the negative and positive energy states while the other three modes are gapped.
The vacuum state is obtained by filling the negative energy states, corresponding to the Dirac sea.
We can expand the resulting fermion wave function in the energy eigenbasis as 
\begin{align}
	\psi_\text{L} = &\,
	e^{i \left( k + \frac{f}{2} \right) \eta} d{^{(+1)\,\dag}_{\text{L}, - \bm{k}}} e_{\bm k}^{(+1)}
	+  \left[
	e^{-i \left( k - \frac{f}{2} \right) \eta}\, \theta \left(\frac{f}{2} - k\right) d{^{(-1)\,\dag}_{\text{L},- \bm{k}}}
	+  e^{-i \left( k - \frac{f}{2} \right) \eta}\, \theta \left(k - \frac{f}{2} \right) b{^{(-1)}_{\text{L},\bm{k}}}
	\right]
	e_{\bm k}^{(-1)} \nonumber \\[.5em]
	& \qquad + e^{i \left( \sqrt{k^2 + f^2} - \frac{f}{2} \right) \eta} d{^{(0)\,\dag}_{\text{L},- \bm{k}}} e_{\bm k}^{(0;1)} 
	+ e^{- i \left( \sqrt{k^2 + f^2} + \frac{f}{2} \right) \eta} b^{(0)}_{\text{L},\bm{k}} e^{(0;2)}_{\bm k}\,,
	\label{eq:mode_const_f}
\end{align}
with the vacuum being annihilated by $d^{(\bullet)}_{\text{L},\bm k} \ket{0} = b^{(\bullet)}_{\text{L},\bm k} \ket{0} = 0$ (with $d$ associated with negative energy states and $b$ with positive energy states). {Here we have adopted the following normalization for the commutators:
$\{b_{\text{L},\bm k}^{(p)}, b_{\text{L},\bm q}^{(p')}{}^\dag \} = (2 \pi)^3 \delta_{pp'} \delta (\bm{k} - \bm{q})$ and $\{ d_{\text{L},\bm k}^{(n)}, d_{\text{L},\bm q}^{(n')}{}^\dag \} = (2 \pi)^3 \delta_{nn'} \delta (\bm{k} - \bm{q})$.}
{Note that the mixed modes $\psi_\text{L/R}^{(0;1)}$ and $\psi_\text{L/R}^{(0;2)}$ always feature exactly one positive and one negative frequency mode per helicity, so that the corresponding annihilation operators can be simply denoted by $b^{(0)}$ and $d^{(0)}$, respectively. Along the same lines, note}
{that the Heaviside theta function must appear together with $d_{\text{L}, - \bm k}^{(-1)}$ and $b_{\text{L},\bm k}^{(-1)}$ since they are only defined for $k < f / 2$ and $k > f / 2$ respectively.}
{
In the rest of this paper we usually omit this theta function for a notational simplicity unless it leads to some confusions.
}
The eigenvectors $e^{(\bullet)}_{\bm k}$ introduced in Eq.~\eqref{eq:mode_const_f} are constructed to diagonalize the Hamiltonian,
\begin{align}
	e^{(+1)}_{\bm k} \equiv \chi^{(+)}_{\bm k} t^{(+)}_{\bm k}\,,
	\quad
	e^{(-1)}_{\bm k} \equiv \chi^{(-)}_{\bm k} t^{(-)}_{\bm k}\,,
	\quad
	\left(\begin{array}{c}
	e_{\bm k}^{(0;1)}\\
	e_{\bm k}^{(0;2)} \end{array}\right) = 
	O(k/f)\,
	\left(\begin{array}{c}
	\chi^{(+)}_{\bm k} t^{(-)}_{\bm k} \\
	\chi^{(-)}_{\bm k} t^{(+)}_{\bm k} \end{array}\right)\,.
	\label{eq:basis}
\end{align}
Note that the dispersion relations for the left- and right-handed fermion interchange (after re-labeling the states) under flipping the sign of $f$, \textit{i.e.}, $\omega_\text{L} \leftrightarrow \omega_\text{R}$ for $f \leftrightarrow - f$.
This behavior is expected because the sign-flipping $f \mapsto -f$ is nothing but the parity transformation as can be seen from Eq.~\eqref{eq:gauge_bkg}. In other words, once we have solved the equation of motion for, \textit{e.g.}, all left-handed fermions on a given background $f$, we can obtain the solution for the right-handed fermions by simply flipping the sign of $f$, since the spectrum of left-handed fermions in a background $-f$ is equivalent to the spectrum of right-fermions in the background $+f$.

\paragraph{Strategy to analyze particle production.}
Now we are in a position to discuss how to estimate particle production induced by the evolution of $f (\eta)$.
To define the notion of particle and anti-particle unambiguously, we assume that $f (\eta)$ takes different constant values in the far past and far future, \textit{i.e.},
\begin{align}
	f (\eta) &= 
	\begin{cases}
	f_i & \text{for}\quad \eta \leq \eta_i\,,\\
	f_f & \text{for}\quad \eta_f \leq \eta\,.
	\end{cases}
	\label{eq:f_evolve}
\end{align}
We also assume $0 < f_i < f_f$ for simplicity.
For $\eta \leq \eta_i$ and $\eta_f \leq \eta$, one may unambiguously distinguish the positive/negative frequency modes and expand the fermion field as done in Eq.~\eqref{eq:mode_const_f}.

Suppose that we start with the vacuum state which is erased by annihilation operators defined at $\eta \leq \eta_i$.
Let $f(\eta)$ evolve until $f = f_f$ at $\eta_f \leq \eta$.
Then in general the positive (negative) frequency mode defined at $\eta \leq \eta_i$ is no longer purely positive (negative) for $\eta_f \leq \eta$, rather it will contain contributions with positive and negative frequencies.
This leads to the particle production.
There are two mechanisms of fermion production in our setup.
The first is an adiabatic process:
particle production for an arbitrary slow evolution of $f(\eta)$, \textit{i.e.}, 
$|f'/f^2| \to 0$.
To estimate this process, we do not have to specify the evolution of $f(\eta)$ as long as it is slow enough.
We will see in Sec.~\ref{sec:anomaly} that this is related to particle production through the chiral anomaly. Note that the anomaly equation depends only on the difference of $f_i$ and $f_f$ regardless of the details of $f(\eta)$ [see Eq.~\eqref{eq:toshow} below].
The second process is  non-adiabatic: particle production associated with a finiteness of $f'$. In reality, the time derivative of $f$ is finite, \textit{e.g.}, in chromo-natural inflation, we expect 
$|f'/f^2| \sim H/\hat f$.
As an instructive example, we will fix the evolution of $f(\eta)$ as indicated by chromo-natural inflation
and evaluate the associated fermion production in Sec.~\ref{sec:fermion_prod_cni}.

\section{Chiral anomaly and eta invariant}
\label{sec:anomaly}

In this section we compute fermion production in a homogeneous and isotropic gauge field background.
We can compute the fermion production explicitly by solving the equations of motion, given in Eqs.~\eqref{eq:++} -- \eqref{eq:mixed}. We will find the resulting particle production to be asymmetric, $\delta \langle Q_R \rangle = - \delta \langle Q_L \rangle$. This channel of fermion production is sourced by the Chern-Pontryagin density, and must obey the chiral anomaly equation~\eqref{eq:axial},
\begin{align}
	\partial_\eta \vev{Q_\text{L/R}} = \mp \,
	\text{vol}\, (\mathbb{R}^3)\, 
	\frac{1}{8\pi^2} \partial_\eta \left( f^3 \right)
	\quad
	\leftrightarrow
	\quad
	\delta \vev{Q_\text{L/R}} = \mp \,
	\text{vol}\, (\mathbb{R}^3)\,  
	\frac{1}{8\pi^2} \left( f_f^3 - f_i^3 \right) \,,
	\label{eq:toshow}
\end{align}
where we have inserted Eq.~\eqref{eq:topo_dens}. 
Our primary goal of this section is to see by explicit computation that this is indeed the case.
The anomaly equation~\eqref{eq:toshow} 
is not sensitive to how $f(\eta)$ evolves, depending only on the initial and final values $f_i$ and $f_f$. Thus in this section we will assume an adiabatic evolution of $f(\eta)$ to single out the production via the chiral anomaly by suppressing the production related to a finiteness of $f'$ (which will be discussed in Sec.~\ref{sec:app_cni}).

For an arbitrary slow evolution of $f(\eta)$, we cannot create particles on the gapped modes. 
Hence one may concentrate on the gapless mode,
namely the lowest mode $\psi_\text{L/R}^{(-,-)}$ for $f > 0$ (see Fig.~\ref{fig:dispersion}),
and just count the number of states that cross zero energy, 
as was done in the U$(1)$ case in~\cite{Nielsen:1983rb}.
In our case, however, this is not the whole story.
In fact, we will explicitly show that this process explains only $1/6$ of the anomaly equation\footnote{
	For a fundamental representation. See appendix~\ref{sec:general} for a general representation.
} [see Eqs.~\eqref{eq:Qe_L} and \eqref{eq:Qe_R}].
In order to correctly reproduce the anomaly equation, 
we must also take into account a contribution from the vacuum, 
whose meaning we will clarify in the following.
For this purpose, we have to first go back to the definition of the fermion current 
and investigate \textit{how the current must be regularized}.
We will clarify the conditions under which the vacuum contribution is relevant
as well as the essential differences between {the abelian analysis}~\cite{Nielsen:1983rb} and this work in the end of this section.

\subsection{Regularization and eta invariant}
First of all, we discuss a proper regularization of the fermion current.
Generally speaking, the fermion current is divergent, and hence we have to regulate it.
{In particular, the regularization should not spoil the underlying symmetry of the theory.
Here we take the gauge and $CP$ symmetries as our guiding principle.
The $CP$ symmetry is crucial to understand why it is insufficient to count the number of zero-crossing modes
(which corresponds to taking a normal ordering and simply dropping the divergence) to reproduced the anomaly equation.}

\paragraph{$CP$ transformation.}
Since this theory never violates $CP$ explicitly, 
we shall require that the regularization does not spoil the transformation law,
\begin{align}
	CP\, \left[ Q_\text{L/R} (\eta) \right]_{\hat{\Lambda}} \, (CP)^{-1}
	= - \left[ Q_\text{L/R} (\eta)  \right]_{\hat{\Lambda}} \,,
	\label{eq:reg_cp}
\end{align}
where ${\hat{\Lambda}}$ is a \textit{physical} cutoff which is taken to be infinity in the end.
See Eq.~\eqref{eq:current_CP} for the CP transformation of the fermion current.
Anticipating that the $CP$ transformation exchanges the particles and anti-particles, let us redefine the fermion current by combining $\psi$ and $\psi^\dag$ antisymmetrically, \textit{i.e.}, 
$J_\text{L/R}^\mu = [ \overline{\psi}, \mathcal{P}_\text{L/R}\gamma^\mu \psi]/2$.
This redefinition is always possible by adding a total derivative to the action. 
Inserting Eq.~\eqref{eq:mode_const_f}, 
 the associated charge now reads 
\begin{align}
\label{eq:q_antisym}
	\left[ Q_H \right]_{\hat{\Lambda}}
	= \int \frac{\dd^3 k}{(2 \pi)^3}\, \frac{1}{2}
	\Bigg\{ &
	\sum_{p} R \left(\frac{|\omega_H^{(p)}|}{{a \hat{\Lambda}}} \right)
	\left[ b^{(p)\, \dag}_{H,\bm k} b{^{(p)}_{H,\bm k}} 
	- b{^{(p)}_{H,\bm k}} b^{(p)\, \dag}_{H,\bm k}
	\right] \nonumber \\
	& \qquad  -
	\sum_{n} R \left(\frac{|\omega_H^{(n)}|}{a{\hat{\Lambda}}} \right)
	\left[ d^{(n)\, \dag}_{H,- \bm k} d{^{(n)}_{H,- \bm k}} - 
	d{^{(n)}_{H,- \bm k}} d^{(n)\, \dag}_{H,- \bm k}
	\right]
	\Bigg\} \,,
\end{align}
for $H = \text{L}, \text{R}$.
Here the superscript $p$ ($n$) labels the positive (negative) energy mode whose energy is $\omega_H^{(p)}$ ($\omega_H^{(n)}$).
The regulator function $R(x)$ is smooth and rapidly approaches to zero.
Note that the energy spectrum, $\omega_H^{(p)}$ and $\omega_H^{(n)}$, respects the gauge invariance,
and hence so does the regularization $R(|\omega_H^{(\bullet)}|/{a\hat{\Lambda}})$.

Let us see that this regularized current fulfills the required property \eqref{eq:reg_cp}.
{To show this, we first have to understand how the creation and annihilation operators transform under $CP$.
Inserting the mode expansion given in Eq.~\eqref{eq:mode_const_f} into the definition of the $CP$ transformation, $	CP\, \psi_\text{L/R} (x) \, (CP)^{-1} = \mp (i \sigma^2 ) \, T_C \, \psi_\text{L/R}^\dag (x_P)$ with $(x_P) = (\eta, - \bm{x})$, one can obtain these transformation laws.
Note that we are considering here a $CP$ transformation or the fermion $\psi$ on a given background $f$, which spontaneously breaks $CP$-invariance. Effectively, this means that performing a $CP$ transformation amounts to flipping the sign of $f$ in all operators in the fermion equation of motion. In particular, this implies
the following transformation law for the dispersion relation:
\begin{align}
	\omega^{(-1)}_H
	\overset{CP}\longleftrightarrow 
	- \omega^{(+1)}_H \,
	, \quad
	\omega^{(0;1)}_H
	\overset{CP}\longleftrightarrow
	- \omega^{(0;2)}_H \, ,
\end{align}
for $H = \text{L/R}$.
Consequently the $CP$ transformation exchanges the creation/annihilation operator between the particle and anti-particle as follows:
\begin{align}
	&CP\, \left[ \theta \left( k - \frac{f}{2} \right) b_{\text{L},\bm k}^{(-1)}+ \theta \left( \frac{f}{2} - k \right) d_{\text{L},- \bm k}^{(-1)\, \dag}\right] \, (CP)^{-1} 
	= d_{\text{L},- \bm k}^{(+1)}\,, \nonumber \\[.5em]
	&CP\, d_{\text{L},- \bm k}^{(+1)}\, (CP)^{-1} = \theta \left( k - \frac{f}{2} \right) b_{\text{L},\bm k}^{(-1)} + \theta \left( \frac{f}{2} - k \right) d_{\text{L},- \bm k}^{(-1)\, \dag}, \nonumber \\[.5em]
	&CP\, d_{\text{L},- \bm k}^{(0)}\, (CP)^{-1} = b_{\text{L},\bm k}^{(0)}\,, \quad
	CP\, b_{\text{L}, \bm k}^{(0)}\, (CP)^{-1} = d_{\text{L},- \bm k}^{(0)}\,,
\end{align}
for the left-handed fermion; and
\begin{align}
	&CP\, \left[ \theta\left( k - \frac{f}{2} \right) d_{\text{R}, -\bm k}^{(-1)\, \dag} + \theta \left( \frac{f}{2} - k \right) b_{\text{R},\bm k}^{(-1)} \right]  \, (CP)^{-1} = b_{\text{R},\bm k}^{(+1)\, \dag}\,, \quad \nonumber \\[.5em]
	&CP\, b_{\text{R},\bm k}^{(+1)\, \dag}\, (CP)^{-1} = \theta \left( k - \frac{f}{2} \right) d_{\text{R}, - \bm k}^{(-1)\, \dag} + \theta \left( \frac{f}{2} - k \right) b_{\text{R},\bm k}^{(-1)}\, , \nonumber \\[.5em]
	&CP\, b_{\text{R}, \bm k}^{(0)}\, (CP)^{-1} = d_{\text{R}, -\bm k}^{(0)}\,, \quad
	CP\, d_{\text{R}, - \bm k}^{(0)}\, (CP)^{-1} = b_{\text{R}, \bm k}^{(0)}\,,
\end{align}
for the right-handed fermion.
Here we have explicitly written down the Heaviside theta function to avoid confusions.}
By using them we confirm that the regularized current respects the required property~\eqref{eq:reg_cp}.
See appendix~\ref{sec:eta_reg} for details.

\paragraph{Normal ordering and eta invariant.}
The physical meaning of this regularization procedure becomes clear once we rewrite Eq.~\eqref{eq:q_antisym} by singling out the \textit{normal ordering} term by re-writing $b b^\dag = b^\dag b - 1 = \normord{b b^\dag} -1$ (and analogous for the operator $d$),
\begin{align}
	Q_H 
	= \lim_{{\hat{\Lambda}} \to \infty} \left[ Q_H \right]_{\hat{\Lambda}}
	=\normord{Q_H}
	+ \lim_{{\hat{\Lambda}} \to \infty} \text{vol}\, (\mathbb{R}^3) \int \frac{\dd^3 k}{(2 \pi)^3}\,
	\left[ 
		- \frac{1}{2} \sum_\lambda \text{sgn}\, \left( \omega^{(\lambda)}_{H} \right) \,  R \left( \frac{|\omega^{(\lambda)}_H|}{{a \hat{\Lambda}}}\right)
	\right] \,, \label{eq:QH}
\end{align}
where $\text{sgn}\, (x)$ is a sign function and $\lambda = \{(+1), (-1), (0;1), (0;2) \}$ labels the energy eigenstates. 
Here we have written the limit ${\hat{\Lambda}} \to \infty$ explicitly.
The first term, $:Q_H:$, counts the number of particles minus anti-particles as usual.
If we naively apply normal ordering, only this term remains.
The second term is known as the \textit{eta invariant} which measures the contribution from vacuum~\cite{Atiyah:1963zz,Atiyah:1968mp}, {which without regularization is ill-defined}.
It is now clear that we must include the eta invariant in our computation unless the spectrum is identical between the positive and negative energy states, $|\omega_H^{(p)}| = |\omega_H^{(n)}|$. 
Obviously, in our setup, this is not the case (see Fig.~\ref{fig:dispersion}) 
and hence we must keep the eta invariant to reproduce the anomaly equation. 
At the end of this section we discuss in more detail the conditions under which we can ignore the eta invariant.
{Note that Eq.~\eqref{eq:QH} holds for both helicities separately, and the total chiral charge $\bar \psi \gamma^0 \gamma_5 \psi$ is obtained as $Q_\text{R} - Q_\text{L} = 2 Q_\text{R}$.}

\subsection{Chiral asymmetry from excitations}

We first estimate the asymmetry coming from the normal ordering term. 
Assuming an adiabatic evolution of $f(\eta)$ from $f_i$ to $f_f$ [see Eq.~\eqref{eq:f_evolve}], we may count the number of states whose frequency changes from positive to negative or visa versa by just looking at the lowest mode, $\psi_\text{L/R}^{(-1)}e_{\bm k}^{(-1)}$ (for $f>0$).

\paragraph{Solving the equation of motion.}
The equation of motion for $\psi^{(-1)}_\text{L}$, given by
\begin{align}
	0 =  \left[ i \partial_\eta - \left(k - \frac{f(\eta)}{2} \right) \right] \psi_\text{L}^{(-1)} (\eta, \bm{k}),
\end{align}
can be solved formally as
\begin{align}
	\psi^{(-1)}_\text{L} (\eta, {\bm k}) = 
	e^{ - i \int^\eta \dd \bar \eta\, \left( k - \frac{f (\bar \eta)}{2} \right)}\, \left[ \theta \left( k - \frac{f_i}{2} \right) \, b^{(-1)}_{\text{L},\bm k}
	+ \theta \left(\frac{f_i}{2} - k \right) d{^{(-1)\,\dag}_{\text{L},- \bm k}} \right] \,.
\end{align}
Here we take the initial condition to match the vacuum state erased by the annihilation operators defined at $\eta \leq \eta_i$, \textit{i.e.}, $\vev{:Q_H:} = 0$ for $\eta \leq \eta_i$.
In other words, this solution becomes
\begin{align}
	\psi_\text{L}^{(-1)}
	= e^{ - i \left( k - \frac{f_i}{2} \right)\eta}\, \theta \left( k - \frac{f_i}{2} \right) \, b^{(-1)}_{\text{L},\bm k}
	+ \theta \left(\frac{f_i}{2} - k \right) \, e^{i \left( \frac{f_i}{2} - k \right) \eta} d{^{(-1)\,\dag}_{\text{L},- \bm k}} \,,
\end{align}
for $\eta \leq \eta_i$. Eventually, for $\eta_f \leq \eta$, this reads 
\begin{align}
	\psi_\text{L}^{(-1)}
	= e^{ - i \left( k - \frac{f_f}{2} \right)\left( \eta - \eta_f \right)}
	e^{- i \delta}
	\Bigg\{
	\theta \left( k - \frac{f_f}{2} \right) \,  \underbrace{b^{(-1)}_{\text{L},\bm k}}_{B^{(-1)}_{\text{L},\bm k}}
	+ \theta \left(\frac{f_f}{2} - k \right) \, 
	\underbrace{
	\left[\theta \left( k - 
	\frac{f_i}{2} 
	\right) b^{(-1)}_{\text{L},\bm k} + d{^{(-1)\,\dag}_{\text{L},- \bm k}} \right]}_{D{^{(-1)\,\dag}_{\text{L},- \bm k}}} \Bigg\} \,,
	\label{eq:-1_f}
\end{align}
where $\delta \equiv \int^{\eta_f}_{\eta_i} \dd \bar \eta\, \left( k - f (\bar \eta)/2 \right)$.
Here $B_{\text{L},\bm k}^{(\dag)}$ and $D_{\text{L},- \bm k}^{(\dag)}$ are annihilation (creation) operators defined at $\eta_f \leq \eta$. They cannot erase the initial state, which implies particle or anti-particle production.
This solution can easily be understood from Fig.~\ref{fig:dispersion}.
When $f$ grows, some of the positive energy states, which are vacant, becomes negative energy states.
This is why the creation operator of anti-particles, $D^\dag$, contains the annihilation operator of particles, $b$, of the initial state.

\paragraph{Chiral asymmetry.}
Taking the expectation value of $:Q_L:$ in the initial vacuum state, we arrive at
\begin{align}
	\delta Q_\text{L}^\text{(e)} 
	&\equiv \vev{ \normord{Q_\text{L} (\eta_f)}} - \vev{\normord{Q_\text{L} (\eta_i)}} \\[.5em]
	& = - \text{vol}\, (\mathbb{R}^3) \int \frac{\dd^3 k}{(2 \pi)^3}\, 
	\theta \left( k -  
	\frac{f_i}{2} 
	\right) \theta \left( 
	\frac{f_f}{2}
	 - k \right) \\[.5em]
	&= \frac{1}{6} \times \left[ - \text{vol}\, (\mathbb{R}^3)\, 
	\frac{1}{8 \pi^2}
	 \left( f_f^3 - f_i^3 \right)
	\right] \,.
	\label{eq:Qe_L}
\end{align}
{We note that in this computation, $\delta Q_L^\text{(e)}$ only depends on the zero of the dispersion relation, as well as on the change in the background gauge field, $f_f^3 - f_i^3$.} 
A similar computation yields the asymmetry of right-handed fermion as
\begin{align}
	\delta Q_\text{R}^\text{(e)} = \frac{1}{6} \times \left[ \text{vol}\, (\mathbb{R}^3)\,
	\frac{1}{8 \pi^2} 
	\left( f_f^3 - f_i^3 \right)
	\right] \,.
	\label{eq:Qe_R}
\end{align}

It is instructive to reproduce the sign of Eqs.~\eqref{eq:Qe_L} and \eqref{eq:Qe_R} from Fig.~\ref{fig:dispersion}.
As can be seen from the left panel of Fig.~\ref{fig:dispersion},
if $f$ grows (note that $f > 0$), 
the positive energy states, which are vacant, become the negative energy states.
Thus, anti-particles are generated for the left-handed fermion, \textit{i.e.}, $\delta Q^\text{(e)}_\text{L} < 0$.
For the right-handed fermion, on the other hand,
the negative energy states, which are occupied, turn into the positive energy states. Namely, particle production occurs for the right-handed fermion, \textit{i.e.}, $\delta Q^\text{(e)}_\text{R} > 0$.

By comparing Eqs.~\eqref{eq:Qe_L} and \eqref{eq:Qe_R} with Eq.~\eqref{eq:toshow}, we conclude that the particle or anti-particle excitations account for just $1/6$ of the asymmetry predicted by the anomaly equation.
If we take a spin-$j$ representation of SU$(2)$ in stead, 
the factor $1/6$ becomes $j^3 / (j+1) (j + 1/2) j$.\footnote{
	{In the large $j$ limit,  the contribution only from excitations suffices to recover the anomaly equation.
	This is because the spectrum becomes symmetric and hence we can ignore the eta invariant in this limit.
	See the discussion in the end of this section and appendix~\ref{sec:general}.}
} See appendix~\ref{sec:general} for details.
For later convenience, we rewrite Eqs.~\eqref{eq:Qe_L} and \eqref{eq:Qe_R} in terms of number density. The number density generated via the chiral anomaly can be expressed as 
\begin{align}
	\left. \bar n_\text{L} \right|_\text{adiabatic} = \left. n_\text{R} \right|_\text{adiabatic} = \frac{1}{6}  
	\frac{1}{8 \pi^2} 
	(f_f^3 - f_i^3) \,, \quad
	\left. n_\text{L} \right|_\text{adiabatic} = \left. \bar n_\text{R} \right|_\text{adiabatic} = 0 \,.
	\label{eq:num_anomaly}
\end{align}

\subsection{Chiral asymmetry from the vacuum}

We now move on to the asymmetry from the vacuum contribution.

\paragraph{Chiral asymmetry.}
The vacuum contribution is defined by
\begin{align}
	Q_H^\text{(v)} \equiv \lim_{{\hat{\Lambda}} \to \infty} \text{vol}\, (\mathbb{R}^3) \int \frac{\dd^3 k}{(2\pi)^3}\,
	\left[ 
		- \frac{1}{2} \sum_\lambda \text{sgn}\, \left( \omega^{(\lambda)}_{H} \right) \,  R \left( \frac{ | \omega^{(\lambda)}_H |}{{a \hat{\Lambda}}}\right)
	\right] \,.
	\label{eq:eta_inv_def}
\end{align}
First, suppose that $f(\eta)$ takes a constant value. 
Assuming for instance a Gaussian regulator, $R(x) = e^{-x^2}$, one may compute $Q_H^\text{(v)}$ explicitly by plugging in the dispersion relations given in Eqs.~\eqref{eq:disp_++}, \eqref{eq:disp_--}, \eqref{eq:disp_01}, and \eqref{eq:disp_02}.
Moreover, as expected, we can show that the result is independent of a concrete form of the regulator if it fulfills appropriate properties.
See appendix~\ref{sec:indep} for an explicit proof.
A straightforward calculation leads to
\begin{align}
	Q_\text{L/R}^\text{(v)} = \frac{5}{6} \times \left[
	\mp \text{vol}\, (\mathbb{R}^3)\, 
	\frac{1}{8\pi^2} f^3
	\right] \,.
	\label{eq:Qv_const}
\end{align}

Next we consider the evolution given in Eq.~\eqref{eq:f_evolve}.
In the end, we are interested in a difference of $Q_H^\text{(v)}$ between its initial and final value.
Since Eq.~\eqref{eq:Qv_const} applies to both $\eta \leq \eta_i$ and $\eta_f \leq \eta$, we obtain
\begin{align}
	\delta Q_\text{L/R}^\text{(v)} &\equiv Q_\text{L/R}^\text{(v)} (\eta_f) - Q_\text{L/R}^\text{(v)} (\eta_i) \\[.5em]
	&= 
	\frac{5}{6} \times \left[
	\mp \text{vol}\, (\mathbb{R}^3)\, 
	\frac{1}{8\pi^2} 
	\left(f_f^3 - f_i^3 \right)
	\right].
	\label{eq:Qv}
\end{align}

\paragraph{ }
Summing the contributions from the excitations and vacuum [Eqs.~\eqref{eq:Qe_L}, \eqref{eq:Qe_R} and \eqref{eq:Qv}], we finally correctly reproduce the anomaly equation given in Eq.~\eqref{eq:toshow}:
\begin{align}
	\delta \vev{Q_\text{L/R}} = \delta Q_\text{L/R}^\text{(e)} + \delta Q_\text{L/R}^\text{(v)} 
	= \mp \text{vol}\, (\mathbb{R}^3)\, 
	\frac{1}{8\pi^2} 
	\left(f_f^3 - f_i^3 \right) \,.
	\label{eq:anomaly_rep}
\end{align}
This clearly shows that the vacuum contribution is indispensable. The asymmetric fermion production in a homogeneous isotropic non-abelian gauge field background can thus be understood at the microphysical level as  the sum of two effects: (i) production of the lowest lying, gapless fermion mode due to a time-dependent dispersion relation and (ii) a vacuum contribution due to an asymmetric energy spectrum of all fermion modes.

{
\paragraph{When is the vacuum contribution important?}
In hindsight, it is easy to see the conditions under which we need to take into account the vacuum contribution.
The vacuum contribution is described by the eta invariant:
\begin{align}
	Q_H^\text{(v)} \equiv 
	\lim_{{\hat{\Lambda}} \to \infty} \text{vol}\, (\mathbb{R}^3) \int \frac{\dd^3 k}{(2 \pi)^3}\,
	\left[ 
		- \frac{1}{2} \sum_\lambda \text{sgn}\, \left( \omega^{(\lambda)}_{H} \right) \,  R \left( \frac{|\omega^{(\lambda)}_H|}{{a \hat{\Lambda}}}\right)
	\right]\,.
\end{align}

We can omit the vacuum contribution if one of the following two conditions is fulfilled: 
(1) initial and final gauge configurations are identical,
or (2) positive and negative frequency modes are symmetric. 
Below we briefly explain each of these conditions.  

(1): We are computing a difference between the initial and the final states.
If the initial and final gauge field configurations are equivalent up to gauge transformation,
the initial and final spectra must be the same, resulting in the cancellation of the contribution from the eta invariant.
Indeed this is the case in~\cite{Nielsen:1983rb,Domcke:2018eki}, 
and hence is the reason why they reproduce the anomaly equation
without taking into account the vacuum contribution.
In our case the magnetic fields are different in the initial and final states, 
indicating that they are not connected by the gauge transformation.
(2): If the positive and negative frequency modes are symmetric, 
the eta invariant vanishes by definition.
}

\section{Application to chromo-natural inflation}
\label{sec:app_cni}
{
In the previous section we considered an adiabatic evolution of the gauge field configuration,
and reproduced the anomaly equation correctly by paying special attention to the vacuum contribution.
There we ignored the gapped modes because they are not produced in the adiabatic limit.}
Now we are in a position to discuss non-adiabatic contribution to the fermion production, 
\textit{i.e.}, production of the gapped modes. 
For this purpose we need to specify the evolution of $f(\eta)$. 
We will take $f (\eta) = \hat f/(-H\eta)$
motivated by the chromo-natural inflation (see Sec.~\ref{sec:cni}).
Our goal is to determine how many particles are generated during the evolution of the gauge field background from $f_i$ at $\eta_i$ to $f_f$ at $\eta_f$.

\subsection{Fermion production in chromo-natural inflation}
\label{sec:fermion_prod_cni}

\paragraph{Robustness of the anomaly equation.}
We first confirm that the asymmetric particle production, as described by the anomaly equation, is not  modified by the specific non-adiabatic evolution of $f$.
This discussion also clarifies that the modes $\psi^{(+1)}_\text{L/R}$ and $\psi^{(-1)}_\text{L/R}$ do not participate in any additional fermion production besides the one described in Sec.~\ref{sec:anomaly}.

The equations of motion for the highest and lowest modes in Eqs.~\eqref{eq:++} and \eqref{eq:--} are given by
\begin{align}
	0 &= \left[ i \partial_\eta + \left( \lambda k + \frac{\xi}{2(- \eta)} \right) \right] \psi_\text{L}^{(\lambda)} (\eta, \bm{k})\,,
	\label{eq:eom_pm1}
\end{align}
for $\lambda = +1, -1$ where we have taken $\xi \simeq \hat f/H$ [defined in Eq.~\eqref{eq:eom_f}] to be constant.
Again we discuss only the left-handed fermion to avoid unnecessary complications.
It is clear that the argument presented below holds for the right-handed fermion, too.
Eq.~\eqref{eq:eom_pm1} can be solved analytically, leading to
\begin{align}
	\psi_\text{L}^{(+1)} (\eta, {\bm k}) &= e^{ i k \eta - i \frac{\xi}{2} \ln (- k \eta)  }  d_{\text{L},- \bm k}^{(+1)\, \dag} \,,  \\[.5em]
	\psi_\text{L}^{(-1)} (\eta, {\bm k}) &= e^{ - i k \eta - i \frac{\xi}{2} \ln (- k \eta)  } \left[ \theta \left( k + \frac{\xi}{2 \eta_i} \right) b_{\text{L},\bm k}^{(-1)} + \theta \left(\frac{- \xi}{2 \eta_i} - k \right) d_{\text{L},- \bm k}^{(-1)\, \dag}  \right] \,.
	\label{eq:-1_+1}
\end{align}
Here we have chosen the initial conditions to match the vacuum solution at $\eta_i$.
At $\eta = \eta_f$, we find the positive energy modes for $k > \xi / (-2 \eta_f)$ and negative energy modes for $k < \xi / (-2 \eta_f)$, in other words, part of the initially positive mode has become negative:\footnote{
	{The positive and negative modes at a given time are defined by the Hamiltonian for the fermion
	at that time. Note that it does not contain any time derivatives.}
}
\begin{align}
	\psi_\text{L}^{(+1)} (\eta_f, {\bm k}) &= e^{ i k \eta_f - i \frac{\xi}{2} \ln (- k \eta_f)  }  \underbrace{d_{\text{L},- \bm k}^{(+1)\,\dag}}_{D_{\text{L}-, \bm k}^{(+1)\,\dag}} \,, \label{eq:+1_cni}
	\\[.5em]
	\psi_\text{L}^{(-1)} (\eta_f, {\bm k}) &= e^{ - i k \eta_f - i \frac{\xi}{2} \ln (- k \eta_f)  } 
	\Bigg\{ \theta \left( k + \frac{\xi}{2 \eta_f} \right) \underbrace{b_{\text{L},\bm k}^{(-1)}}_{B_{\text{L},\bm k}^{(-1)}}
	+ \theta \left(\frac{- \xi}{2 \eta_f} - k \right) 
	\underbrace{\left[ \theta \left( k + \frac{\xi}{2 \eta_i} \right) b_{\text{L},\bm k}^{(-1)} + d_{\text{L},- \bm k}^{(-1)\,\dag} \right]}_{D_{\text{L},- \bm k}^{(-1)\,\dag}}  \Bigg\} \,.
	\label{eq:-1_cni}
\end{align}
Eq.~\eqref{eq:-1_cni} is essentially the same as Eq.~\eqref{eq:-1_f}, and hence the computation performed in the previous section holds. Recall that the vacuum contribution to the chiral anomaly is determined by the difference between the final and initial values, independently of the evolution of $f(\eta)$.
Therefore, the chiral anomaly [Eq.~\eqref{eq:anomaly_rep}] is reproduced as expected.
Moreover, Eqs.~\eqref{eq:+1_cni} and \eqref{eq:-1_cni} show that there is no additional fermion production associated with $f' \neq 0$ from $\psi_\text{L}^{(-1)}$ and $\psi_\text{L}^{(+1)}$.

\paragraph{Pair production from $\mathbf{f' \neq 0}$}
In Sec.~\ref{sec:anomaly}, we neglected fermion production from the mixed modes $\psi_\text{L/R}^{(0;1)}$ and $\psi_\text{L/R}^{(0;2)}$ since they are gapped and cannot be generated adiabatically. In the following we extend our discussion to account for effects induced by $f' \neq 0$, focusing on the specific example $f = \hat f / (- H \eta)$. We will see that this leads to fermion production in the mixed modes $\psi_\text{L/R}^{(0;1)}$ and $\psi_\text{L/R}^{(0;2)}$.

Let us go back to the equations of motion given in Eq.~\eqref{eq:mixed}. 
If $f$ is constant, one may diagonalize the equation by the rotation matrix in Eq.~\eqref{eq:rot}.
To single out the effect of non-vanishing $f'$,
it is more convenient to go to this basis.
Then, the equation of motion becomes
\begin{align}
	0 &= 
	\left[
	i \partial_\eta
	 - \left(\begin{array}{cc}
	\omega^{(0;1)}_\text{L/R} (\eta) &    \\
	 &  \omega^{(0;2)}_\text{L/R} (\eta) \end{array}\right) 
	+ \frac{i}{2} \frac{k \xi}{k^2 \eta^2 + \xi^2} \left(\begin{array}{cc}
	& -1   \\
	 1& \end{array}\right) 
	\right]
	\left(\begin{array}{c}
	\psi_\text{L/R}^{(0;1)} (\eta, \bm{k}) \\
	\psi_\text{L/R}^{(0;2)} (\eta, \bm{k}) \end{array}\right) \,.
	\label{eq:eom_mixed}
\end{align}
The third term in the parenthesis encodes the effects coming from $f' \neq 0$. If one drops this term, the results of the previous Sec.~\ref{sec:anomaly} are recovered.
To make this property more explicit, we further define
\begin{align}
\left(\begin{array}{c}
	\psi_\text{L/R}^{(0;1)} (\eta, \bm{k}) \\
	\psi_\text{L/R}^{(0;2)} (\eta, \bm{k}) \end{array}\right)
	\equiv
	\left(\begin{array}{c}
	e^{- i \int^\eta \omega_\text{L/R}^{(0;1)}}\, \varphi_\text{L/R}^{(0;1)} (\eta, \bm{k}) \\
	e^{- i \int^\eta \omega_\text{L/R}^{(0;2)}}\, \varphi_\text{L/R}^{(0;2)} (\eta, \bm{k})
	\end{array}\right) \,.
	\label{eq:ansatz}
\end{align}
In terms of $\varphi_\text{L/R}$, one may easily see the effect of $f' \neq 0$. {For $f' = 0$, inserting Eq.~\eqref{eq:ansatz} in Eq.~\eqref{eq:eom_mixed} leads to a simple plane wave solution with constant amplitude $\varphi$. For $f' \neq 0$, the time evolution of $\varphi$ is determined by}
\begin{align}
	\partial_\eta \varphi_\text{L/R}^{(0;1)} = \frac{1}{2} \frac{k \xi}{k^2 \eta^2 + \xi^2} \, e^{\mp 2 i \Theta} \, \varphi_\text{L/R}^{(0;2)}\,, \quad
	\partial_\eta \varphi_\text{L/R}^{(0;2)} = - \frac{1}{2} \frac{k \xi}{k^2 \eta^2 + \xi^2} \, e^{\pm 2 i \Theta} \, \varphi_\text{L/R}^{(0;1)}\,,
	\label{eq:get_bogo}
\end{align}
where $\Theta \equiv \int^\eta \dd \bar\eta \sqrt{k^2 + \xi^2/\bar \eta^2}$.
Note that this equation is invariant under
\begin{align}
	\varphi_\text{L/R}^{(0;1)} \mapsto - \varphi{_\text{L/R}^{(0;2)}}^{\ast}\,, \quad
	\varphi_\text{L/R}^{(0;2)} \mapsto \varphi{_\text{L/R}^{(0;1)}}^{\ast}\,.
	\label{eq:prop}
\end{align}

We take the left-handed fermion for concreteness in the following.
Pick up one solution $\varphi_{\text{L},+}^{(\bullet)}$ which has a positive frequency initially, \textit{i.e.}, $\varphi_{\text{L},+}^{(0;1)} = 0$ and $\varphi_{\text{L},+}^{(0;2)} = 1$ for $\eta = \eta_i$.
{(Recall that at $\eta_i$, the corresponding frequencies are given by Eqs.~\eqref{eq:disp_01} and \eqref{eq:disp_02}, and for any given helicity, there is always a positive and a negative frequency mode, see Fig.~\ref{fig:dispersion}.)}
Clearly, if we neglect $f'$, $\varphi_{\text{L},+}^{(\bullet)}$ keeps its initial value, \textit{i.e.}, there is no particle production.
Due to the non-vanishing $f'$, a part of the positive mode however turns into the negative one, leading to $\varphi_{\text{L},+}^{(0;1)}(\eta_f) \neq 0$.
We may read off the Bogolyubov coefficients from this solution:\footnote{ {The Bogolyubov transformation is a linear transformation acting on the creation/annihilation operators, $B = \alpha b - \beta^* d^\dagger$ and $D^\dagger = \alpha^* d^\dagger + \beta b$, canonically normalized so that the Bogolyubov coefficients fulfill $|\alpha|^2 + |\beta|^2 = 1 $. 
}}
\begin{align}
	\alpha_{\text{L}, {\bm k}} (\eta_f) \equiv \varphi_{\text{L},+}^{(0;2)} (\eta_f, {\bm k})\,, \quad
	\beta_{\text{L}, {\bm k}} (\eta_f) \equiv \varphi_{\text{L},+}^{(0;1)} (\eta_f, {\bm k})\,.
\end{align}
Instead of taking the positive initial frequency mode, one may consider the negative one, $\varphi^{(\bullet)}_{\text{L},-}$, fulfilling $\varphi_{\text{L},-}^{(0;1)} = 1$ and $\varphi_{\text{L},-}^{(0;2)} = 0$ for $\eta = \eta_i$.
Thanks to Eq.~\eqref{eq:prop}, this solution can also be obtained from $\varphi^{(0;2)}_{\text{L},-} = - \varphi{^{(0;1)}_{\text{L,+}}}^\ast$ and $\varphi^{(0;1)}_{\text{L},-} = \varphi{^{(0;2)}_{\text{L,+}}}^\ast$.
This completes the following Bogolyubov transformation which connects the creation/annihilation operators between $\eta_i$ and $\eta_f$:
\begin{align}
	B_{\text{L},\bm k}^{(0)} = \alpha_{\text{L}, \bm k} b_{\text{L},\bm k}^{(0)} - \beta_{\text{L}, \bm k}^\ast d{_{\text{L}, - \bm k}^{(0)\,\dag}} \,, \quad
	D{_{\text{L},- \bm k}^{(0)\,\dag}} = \beta_{\text{L}, \bm k} b_{\text{L}, \bm k}^{(0)} + \alpha_{\text{L}, \bm k}^\ast d{_{\text{L}, -\bm k}^{(0)\,\dag}} \,.
	\label{eq:bogo}
\end{align}
Because of
$\langle B{_{\text{L}, \bm k}^{(0)}}{}^\dag  B_{\text{L}, \bm k}^{(0)} \rangle = \langle D{_{\text{L}, \bm k}^{(0)}}{}^\dag  D_{\text{L}, \bm k}^{(0)} \rangle = \text{vol}\, (\mathbb{R}^3)\, | \beta_{\text{L}, \bm k} |^2 $,
it describes pair production of particles and anti-particles.
Moreover, one may show that $\alpha_{\text{L},\bm{k}} = \alpha_{\text{R},\bm{k}}$ 
and $\beta_{\text{L},\bm{k}} =  - \beta_{\text{R},\bm{k}}$, meaning that the number of produced particles is identical 
for the left-handed and right-handed fermions, \textit{i.e.}, $|\beta_{\text{L}, \bm k}|^2 = |\beta_{\text{R}, \bm k}|^2$.
Note that the Bogolyubov coefficients transform under $CP$ as
$\alpha \mapsto \alpha$ and $\beta \mapsto - \beta$.

\begin{figure}[t]
	\centering
 	\includegraphics[width=0.5\linewidth]{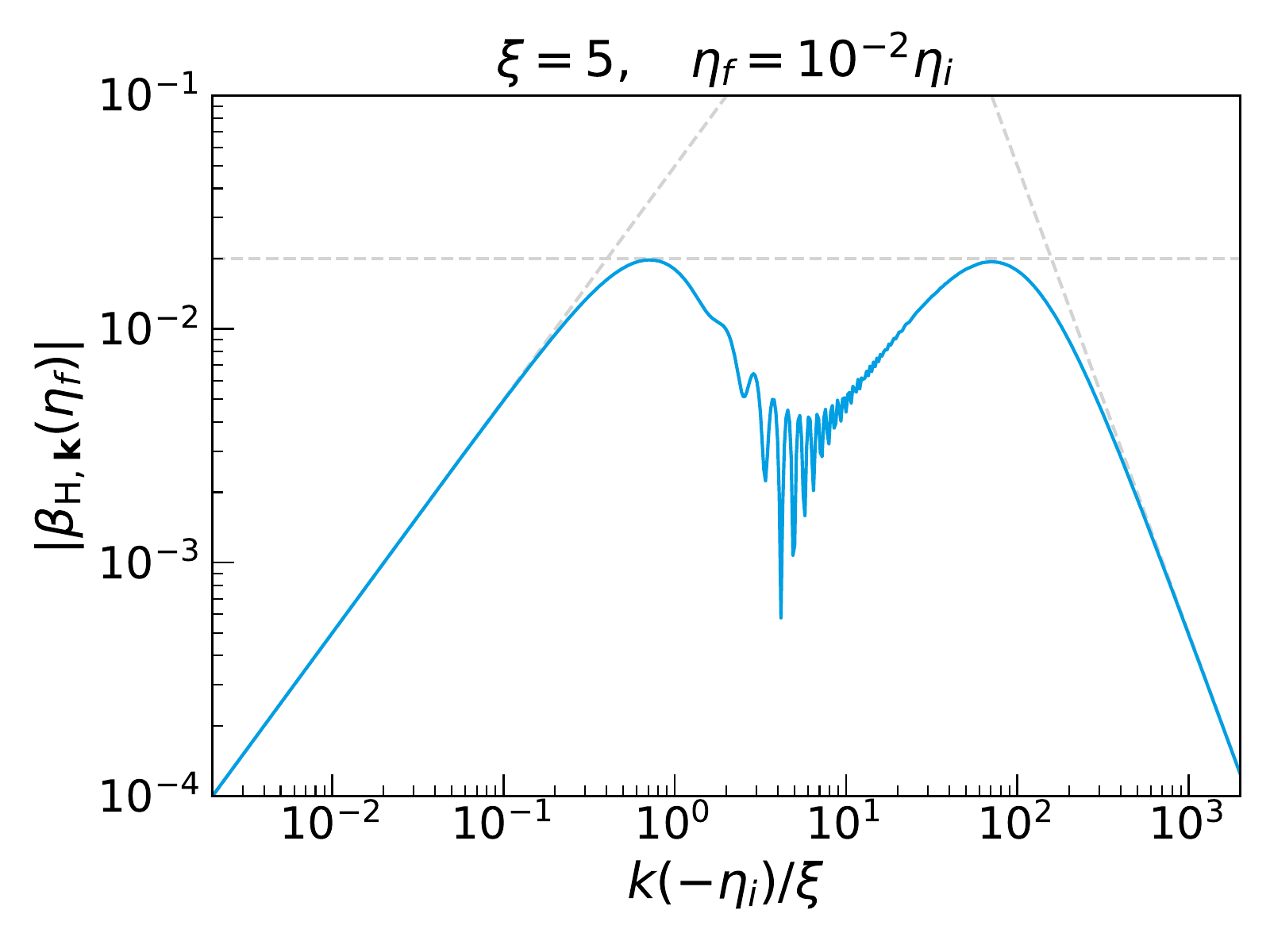}
	\caption{
	The Bogolyubov coefficient $\beta_{\text{H},\bm{k}}$ for $\xi = 5$ and $\eta_f = 10^{-2}\eta_i$.
	The blue line is obtained by numerically solving Eq.~\eqref{eq:get_bogo} without the Born approximation 
	(see appendix~\ref{sec:bogogo}).
	The gray dashed lines are the approximate solution~\eqref{eq:beta_approx}.
	}
	\label{fig:beta}
\end{figure}

One may solve Eq.~\eqref{eq:get_bogo} numerically to get the Bogolyubov coefficients.
Fig.~\ref{fig:beta} shows the numerical result of $|\beta |$ as a function of $\bm k$.
We also display an approximate solution for $|\beta |$ which is given by,
\begin{align}
	\abs{\beta_{\text{H}, \bm k} (\eta_f)} \sim
	\begin{cases}
		\cfrac{k (- \eta_i)}{2 \xi} \cfrac{1}{\sqrt{1 + 4 \xi^2}} & \text{for}~~ k \ll \cfrac{\xi}{- \eta_i} \,, \\[1em]
		\cfrac{c}{4 \xi} & \text{for}~~ \cfrac{\xi}{- \eta_i} \ll k \ll \cfrac{\xi}{- \eta_f} \,, \\[1em]
		\cfrac{\xi}{4 k^2 \eta_f^2} & \text{for}~~ \cfrac{\xi}{- \eta_f} \ll k \,,
	\end{cases}
	\label{eq:beta_approx}
\end{align}
where $H= \text{L}, \text{R}$ and $c \sim 0.4$. The asymptotic behaviour for small and large $k$ is derived in Appendix~\ref{sec:bogogo}, the intermediate range is a rough approximation of the full numerical result with the value of $c$ extracted from the latter.
Using this approximation, we estimate the number density of fermions created by this process.
The comoving number density at $\eta_f$ can be approximated with
\begin{align}
	\left. n_{\text{L}} \right|_\text{pair} = \left. \bar n_{\text{L}} \right|_\text{pair} = \left. n_{\text{R}} \right|_\text{pair} = \left. \bar n_{\text{R}} \right|_\text{pair} =  \int \frac{\dd^3 k}{(2 \pi)^3} \,\abs{\beta_{\text{L/R}, \bm k}}^2
	\sim a^3 \times \frac{1}{24 \pi^2} \xi H^3 \,,
\end{align}
where we have dropped an $\mathcal{O}(1)$ factor arising from the integration of Eq.~\eqref{eq:beta_approx}.
Note that the physical number density is obtained from $\hat n = n / a^3$, and is hence constant {(for constant $\xi$) in a gauge field background described by Eq.~\eqref{eq:gauge_background}}.
Compared to the production via the chiral anomaly [Eq.~\eqref{eq:num_anomaly}], the number density is suppressed by $1/\xi^2$:
\begin{align}
	\left. \bar n_\text{L} \right|_\text{adiabatic} = \left. n_\text{R} \right|_\text{adiabatic} \simeq a^3 \times \frac{1}{4 8 \pi^2} \xi^3 H^3 \,.
\end{align}
Here we have assumed $\eta_i \ll \eta_f\, (<0)$ and denoted $a = a (\eta_f)$.

\subsection{Induced current and backreaction to chromo-natural inflation}
\label{sec:backreaction}
Up to here, we have discussed how the fermions are generated in a gauge field background as found in chromo-natural inflation.
In this section we discuss the backreaction of these produced fermions on the gauge field.

For this purpose, let us go back to the equation of motion for the background gauge field.
Expanding the gauge field as $A^{ai} = f \delta^{ai} + \delta A^{ai}$, one may write down an effective action for the homogeneous gauge field, $f$.
The interaction with the fermion is imprinted in the current, \textit{i.e.}, 
$\overline{\psi} \slashed{A}^a T^a \psi \supset f (\psi^\dag_\text{L} \bm{\sigma} \cdot \bm{T} \psi_\text{L} - \psi^\dag_\text{R} \bm{\sigma} \cdot \bm{T} \psi_\text{R})$.
Differentiating the action with respect to $f$, one may derive the equation of motion including the backreaction from the current:
\begin{align}
	f''(\eta) + 2 f^3(\eta) - \frac{2 \xi}{(- \eta)} f^2(\eta)
	=
	\frac{g^2}{3\, \text{vol}\, (\mathbb{R}^3)}\,
	\int \dd^3 x\,
	\left( \vev{\psi^\dag_\text{L} \bm{\sigma} \cdot \bm{T} \psi_\text{L}} - \vev{\psi^\dag_\text{R} \bm{\sigma} \cdot \bm{T} \psi_\text{R}} \right) \,.
\end{align}
The expectation value of the currents on the right-hand side contains two contributions.
Firstly, there are vacuum fluctuations.
Since a non-vanishing field value of $f$ changes the dispersion relation of fermions, it affects the running of the gauge coupling,
analogous to the Coleman-Weinberg correction to the effective potential~\cite{Coleman:1973jx}.
The second contribution is due to fermions produced from the background gauge field. Once produced they move in the background gauge field, leading to an induced current. 
We first estimate these contributions to the gauge field equation of motion, and then discuss their implications.

\paragraph{Regularization of the current.}
Since the current which couples to the gauge field diverges, we have to regulate it, similar to the situation encountered in Sec.~\ref{sec:anomaly}.
The operator must be regularized in a way that does not spoil the symmetry of the theory, in particular the current must be $CP$ odd as can be seen from $CP: f \mapsto -f$.
For this purpose, analogous to the discussion given in Sec.\ \ref{sec:anomaly}, we redefine the current by antisymmetrizing it, \textit{i.e.},\footnote{{Note that we have used the special structure of the background gauge field to contract gauge and spatial indices here. In a slight abuse of notation, we refer to the resulting scalar object as `current'.}} 
\begin{align}
	\K_\text{L/R} \equiv \pm \int \dd^3 x\, \frac{1}{2} [{\psi}_\text{L/R}^\dag, \bm{\sigma} \cdot \bm{T}\psi_\text{L/R}] \,.
	\label{eq:K}
\end{align}

Let us first write down the current at $\eta > \eta_i$ in terms of creation/annihilation operators defined at $\eta$, \textit{i.e.}, $B$ and $D$. For simplicity, we write down the current of the left-handed fermion only. It is straightforward to obtain its counterpart for the right-handed fermion. {Introducing a regulator function $R$ such that the regularized currents remains $CP$-odd, }
the regularized current can be expressed as
\begin{align}
	\left[ \K_\text{L} \right]_{\hat{\Lambda}}
	\equiv 
	\int \frac{\dd^3 k}{( 2 \pi)^3}\, \Bigg\{ &
		- R \left( \frac{|\omega^{(+1)}_\text{L}|}{{a\hat{\Lambda}}} \right) \frac{1}{2} \left[ D_{\text{L},- \bm k}^{(+1)\, \dag} D_{\text{L},- \bm k}^{(+1)} - \frac{1}{2} \text{vol}\, (\mathbb{R}^3) \right] \nonumber \\[.5em]
		& + R \left( \frac{| \omega^{(-1)}_\text{L} |}{{a\hat{\Lambda}}} \right) \frac{1}{2} \left[ B_{\text{L}, \bm k}^{(-1)\, \dag} B_{\text{L}, \bm k}^{(-1)} - D_{\text{L},- \bm k}^{(-1)\, \dag} D_{\text{L},- \bm k}^{(-1)} - \text{sgn}\, \left(\omega^{(-1)}_\text{L} \right)\, \frac{1}{2} \text{vol}\, (\mathbb{R}^3) \right] \nonumber\\[.5em]
		 &- R \left( \frac{|\omega^{(0;1)}_\text{L}|}{{a\hat{\Lambda}}} \right) 
		 \left( \frac{f}{\sqrt{k^2 + f^2}} - \frac{1}{2} \right)
		 \left[
		 	D^{(0)\,\dag}_{\text{L}, - \bm k} D^{(0)}_{\text{L},- \bm k} - \frac{1}{2} \text{vol}\, (\mathbb{R}^3)
		 \right] \nonumber\\[.5em]
		 &+ R \left( \frac{|\omega^{(0;2)}_\text{L}|}{{a\hat{\Lambda}}} \right) 
	\left( \frac{- f}{\sqrt{k^2 + f^2}} - \frac{1}{2} \right)
		 \left[
		 	B^{(0)\,\dag}_{\text{L}, \bm k} B^{(0)}_{\text{L},\bm k} - \frac{1}{2} \text{vol}\, (\mathbb{R}^3)
		 \right] \nonumber\\[.5em]
		 &+ \frac{k}{2 \sqrt{k^2 + f^2}} 
		 \left[ R \left( \frac{|\omega^{(0;1)}_\text{L}|}{{a\hat{\Lambda}}} \right) + R \left( \frac{|\omega^{(0;2)}_\text{L}|}{{a\hat{\Lambda}}} \right) \right] \left( e^{2 i \Theta} B^{(0)\,\dag}_{\text{L},\bm k} D^{(0)\, \dag}_{\text{L}, - \bm k} + \text{H.c.} \right)
	\Bigg\} \,.
\end{align}
Here we have used the fact that $\bm{\sigma} \cdot \bm{T}$ takes the following form in the basis which diagonalizes the Hamiltonian for the fermion:
\begin{align}
	( \widetilde{ \bm{\sigma} \cdot \bm{T}} ) = 
	\begin{pmatrix}
	\frac{1}{2} &  &  &  \\
	& \frac{1}{2} &  &  \\
	 &  & 
	 \frac{f}{\sqrt{k^2 + f^2}} - \frac{1}{2} & 
	\frac{k}{\sqrt{k^2 + f^2}}  \\
	&  &  
	\frac{k}{\sqrt{k^2 + f^2}}& 
	\frac{- f}{\sqrt{k^2 + f^2}} - \frac{1}{2} 
	\end{pmatrix}
	\begin{matrix}
		\cdots (+1)\\ \cdots (-1)\\ \cdots (0;1) \\ \cdots (0;2)
	\end{matrix}
	\,\,\,\,\, .
\end{align}
Here the tilde denotes that we have taken the energy eigenbasis for the fermion.
Recalling that the $CP$ transformation exchanges the creation/annihilation operators [see \eqref{eq:cp_cran}], the positive/negative energies [see \eqref{eq:cp_energy}], and changes the sign of the gauge field $CP: f \mapsto -f$,
one can show explicitly that the regularized current is $CP$ odd, \textit{i.e.}, $CP: [\K_\text{L/R}]_{\hat{\Lambda}} \mapsto - [\K_\text{L/R}]_{\hat{\Lambda}}$.

To clarify the physical meaning, we divide this regularized current into contributions from vacuum and excitations, as in Sec.~\ref{sec:anomaly}.
One may factor out the {normal ordering} term as follows:
\begin{align}
	\K_H = \lim_{{\hat{\Lambda}} \to \infty} \left[ \K_H \right]_{\hat{\Lambda}}
	= \normord{\K_H} 
	+ \lim_{{\hat{\Lambda}} \to \infty}  [ \K_H^\text{(v)} ]_{\hat{\Lambda}} \,,
	\label{eq:current0}
\end{align}
with
\begin{align}
	[ \K_H^\text{(v)}]_{\hat{\Lambda}} \equiv
	\epsilon_H \,\text{vol}\, (\mathbb{R}^3)
	\int \frac{\dd^3 k}{(2 \pi)^3} \, \Bigg\{ &
	\left[  
		- \frac{1}{2} \sum_\lambda\, (\widetilde{\bm{\sigma} \cdot \bm{T}})_{\lambda \lambda} \, \text{sgn}\, \left(\omega^{(\lambda)}_H \right)\, R \left( \frac{| \omega^{(\lambda)}_H |}{{a\hat{\Lambda}}} \right)
	\right] \label{eq:log_div_pot} \\[.5em]
	& + \frac{k}{2 \sqrt{k^2 + f^2}} 
		 \left[ R \left( \frac{|\omega^{(0;1)}_H|}{{a\hat{\Lambda}}} \right) + R \left( \frac{|\omega^{(0;2)}_H|}{{a\hat{\Lambda}}} \right) \right] \left( - e^{2 i \Theta} \alpha^\ast_{H,\bm k} \beta_{H, \bm k} + \text{H.c.} \right)
		 \Bigg\}\,, \label{eq:log_div_kin}
\end{align}
where we define $\epsilon_H = \pm$ for $H = \text{L/R}$.
Here we have used the Bogolyubov coefficients defined in Eq.~\eqref{eq:bogo} and $\Theta$ introduced below Eq.~\eqref{eq:get_bogo}.
The first term {in Eq.~\eqref{eq:current0}} counts the contribution from particles and anti-particles while the second term stems from vacuum fluctuations.
Contrary to the case of the chiral anomaly, the vacuum contribution actually diverges. However we will see that this divergence is renormalized by the running of the gauge coupling, which eventually leads to the \textit{Coleman-Weinberg-type correction} to the effective potential for $f$.

\paragraph{Vacuum contribution to the current.}
 Here we estimate the vacuum contribution $[ \K_H^\text{(v)}]_{\hat{\Lambda}}$, breaking it down further into two terms to clarify its origin. On the one hand, Eq.~\eqref{eq:log_div_pot} arises due to the asymmetric fermion energy levels, analogous to~\eqref{eq:eta_inv_def}. Assuming for instance a Gaussian regulator, one may compute this integral explicitly, finding a logarithmically divergent term as ${\hat{\Lambda}} \to \infty$.
On the other hand, Eq.~\eqref{eq:log_div_kin} becomes non-zero only if we take into account the time evolution of $f$, since the Bogolyubov coefficient, $\beta$, vanishes for $f = \text{const}$, see Sec.~\ref{sec:fermion_prod_cni}. To compute this contribution, 
we need to evaluate $\alpha$ and $\beta$ for Eq.~\eqref{eq:log_div_kin}.
Since we are interested in its divergence structure, we can take the large momentum limit, \textit{i.e.}, $k (- \eta) \gg \xi$.
We obtain in this limit (see appendix~\ref{sec:bogogo})
\begin{align}
	\alpha_{\text{L}, \bm k} \simeq 1\,, \quad
	\beta_{\text{L}, \bm k} \simeq \frac{\xi e^{- 2 i k \eta}}{4} \left( \frac{i}{k^2 \eta^2} - \frac{1}{k^3 \eta^3} \right) \,, \quad
	\Theta \simeq k \eta \,.
	\label{eq:beta_nlo}
\end{align}
Inserting this into Eq.~\eqref{eq:log_div_kin}, one immediately finds a logarithmic divergence.
Summing the logarithmically  divergent terms coming from Eq.~\eqref{eq:log_div_pot} and \eqref{eq:log_div_kin}, we find
\begin{align}
	[ \K_\text{L}^\text{(v)} ]_{\hat{\Lambda}} \simeq
	\text{vol}\, (\mathbb{R}^3) \times 
	\frac{1}{16 \pi^2} \ln \left( \frac{\hat{f}^2}{{\hat{\Lambda}}^2} \right)\, \left( f'' + 2 f^3 \right)\,,
	\label{eq:vac_op}
\end{align}
{with the term proportional to $f^3$ arising from Eq.~\eqref{eq:log_div_pot} while the $f''$ term arises from Eq.~\eqref{eq:log_div_kin}.}
Here note that $\hat f = f/a$.
The corresponding right-handed current can be obtained from replacing $f$ with $-f$.
We would like to emphasize that this expression \eqref{eq:vac_op} does not depend on $\xi$.
In fact, one may derive the same equation for $f (\eta) = c / (- \eta)$ with $c$ being an arbitrary constant. 
Moreover, we expect this form must hold for a more general evolution of $f (\eta)$ because it is related to the renormalization of the gauge coupling as we discuss below.

Let us show that this divergence can be renormalized by the gauge coupling.
Inserting Eq.~\eqref{eq:vac_op} in the equation of motion for the gauge field, one obtains 
\begin{align}
	0  = - \left[ \frac{1}{g_{\hat{\Lambda}}^2} - \frac{N_F}{48 \pi^2} \ln \left( \frac{\hat{f}^2}{\hat{\Lambda}^2} \right) \right]  \left( f'' + 2 f^3 \right) + \frac{a \dot \phi}{4 \pi^2 f_a} f^2 + \frac{1}{3\, \text{vol}\, (\mathbb{R}^3)}  \sum_H \epsilon_H \vev{ \normord{\K_H}} \,,
	\label{eq:renorm_eom}
\end{align}
with $g_{\hat{\Lambda}}$ being a bare coupling.
$N_F$ counts the number of Weyl fermions and hence $N_F = 2$ in our case 
{(after including the right-handed fermion as well)}.
Recall that the running of the gauge coupling can be expressed as
\begin{align}
	 \frac{1}{g_{\hat\mu}^2} = \frac{1}{g_{\hat\Lambda}^2} + \left[ \frac{22}{3} - \frac{2 N_F}{3} T(\textbf{r}) \right] \frac{1}{16 \pi^2} \ln \left( \frac{\hat{\mu}^2}{\hat{\Lambda}^2} \right)\,.
	 \label{eq:backreaction0}
\end{align}
The first factor of $22/3$ comes from the gauge boson loop.
Since we are interested in the UV-divergence induced by the fermions, we may concentrate on the second term.\footnote{{Perturbations around the homogeneous gauge field background of CNI are studied in~\cite{Dimastrogiovanni:2012st,Dimastrogiovanni:2012ew,Adshead:2013qp,Adshead:2013nka,Domcke:2018rvv}. The most significant effects arise from a tachyonic tensor mode, which receives a temporary exponential enhancement in the gaue field background~\eqref{eq:gauge_bkg_hom}.}}
It is clear that, for $T(\textbf{2}) = 1/2$, the divergence in $g_{\hat\Lambda}$ and $\ln (\hat{f}^2 / \hat{\Lambda}^2)$ cancels out, resulting in the usual logarithmic dependence on the renormalization scale $\hat \mu$.
Though we do not explicitly compute the gauge boson loop in this paper because our main focus is on the contribution from fermions,
this suggests that the full one-loop result is obtained after the inclusion of the gauge boson loop.
Namely one may replace the bare coupling in Eq.~\eqref{eq:backreaction0} by the running gauge coupling,
\begin{align}
	\frac{1}{g_{\hat{\Lambda}}^2} + \left[ \frac{22}{3} - \frac{2 N_F}{3} T(\textbf{r}) \right] \frac{1}{16 \pi^2} \ln \left( \frac{\hat{f}^2}{\hat{\Lambda}^2} \right)
	=
	\frac{1}{g_{\hat{\mu}}^2} + \left[ \frac{22}{3} - \frac{2 N_F}{3} T(\textbf{r}) \right] \frac{1}{16 \pi^2} \ln \left( \frac{\hat{f}^2}{\hat{\mu}^2} \right)\,.
\end{align}
This shows that perturbation theory is under control if we use the running coupling evaluated at $\hat \mu \simeq |\hat f|$.
As a result, we arrive at the following equation of motion:
\begin{align}
	0  = - \frac{1}{g_{\hat f}^2}\left( f'' + 2 f^3 \right) + \frac{a \dot \phi}{ 4 \pi^2 f_a } f^2 + \frac{1}{3\, \text{vol}\, (\mathbb{R}^3)}  \sum_H \epsilon_H \vev{ \normord{\K_H}} \,.
	\label{eq:eom_w_current}
\end{align}
where $g_{\hat f}$ denotes the running gauge coupling evaluated at $\hat{\mu} \simeq |\hat{f}|$.
At this stage one can see the reason why we only have logarithmic divergences and why we can nicely combine them as Eq.~\eqref{eq:vac_op}:
the gauge symmetry restricts the structure of divergence so that it can be renormalized solely by the gauge coupling.

\paragraph{Induced current.}
Now we are in a position to discuss the induced current from the fermionic particles/anti-particles generated from the background gauge field.
The induced current for the left-handed fermion is given by
\begin{align}
	\vev{\normord{\K_\text{L}}} = 
	\int \frac{\dd^3 k}{(2 \pi )^3} \, \Bigg[ &
		\frac{1}{2} \vev{ B^{(+1)\,\dag}_{\text{L}, \bm k} B^{(+1)}_{\text{L}, \bm k} + B^{(-1)\,\dag}_{\text{L}, \bm k} B^{(-1)}_{\text{L}, \bm k} - D^{(-1)\,\dag}_{\text{L}, \bm k} D^{(-1)}_{\text{L}, \bm k} } \label{eq:anomaly_induced}\\[.5em]
	& \qquad
	- \left( \frac{f}{\sqrt{k^2 + f^2}} - \frac{1}{2} \right)
		 	\vev{D^{(0)\,\dag}_{\text{L}, \bm k} D^{(0)}_{\text{L}, \bm k}}
	+ \left( \frac{-f}{\sqrt{k^2 + f^2}} - \frac{1}{2} \right)
	\vev{B^{(0)\,\dag}_{\text{L}, \bm k} B^{(0)}_{\text{L}, \bm k}}
	\Bigg]\,. \label{eq:pair_induced}
\end{align}
One may evaluate the expectation values by using the relation between the creation/annihilation operators defined at $\eta$ and those defined at $\eta = \eta_i$, which is given in Eqs.~\eqref{eq:+1_cni}, \eqref{eq:-1_cni}, and \eqref{eq:bogo}.
The first line [Eq.~\eqref{eq:anomaly_induced}] counts the contribution from the chiral anomaly which gives
\begin{align}
	\vev{ \normord{\K_\text{L}} }_\text{adiabatic}
	= - \text{vol}\, (\mathbb{R}^3) \,
	\frac{1}{96 \pi^2} f^3\,.
	\label{eq:anomaly_induced_result}
\end{align}
Here we take $f_i = f(\eta_i) \to 0$. For right-handed fermions, one may obtain it from just replacing $f$ with $-f$ in Eq.~\eqref{eq:anomaly_induced_result}.
The second line gives the induced current from pair production:
\begin{align}
	\vev{ \normord{\K_\text{L}} }_\text{pair} = 
	 - \int \frac{\dd^3 k}{( 2\pi )^3}\, \frac{2 f}{\sqrt{k^2 + f^2}}\, \abs{\beta_{\text{L}, \bm k}}^2
	 \simeq - \text{vol}\, (\mathbb{R}^3)\, \frac{\tilde c}{96 \pi^2} f''\,,
 	\label{eq:pair_induced_result}
\end{align}
where numerically find that $\tilde c \simeq 0.6$.
The result for right-handed fermions is obtained from just replacing $f$ with $- f$.
Note again that the expression \eqref{eq:pair_induced_result} does not depend on $\xi$, \textit{i.e.}, the same equation can be derived for $f (\eta) = c / (- \eta)$ with $c$ being an arbitrary constant. 
Contrary to the vacuum contribution, we do have to assume the specific time evolution, $f (\eta) \propto 1 / (- \eta)$, which is justified a posteriori as we will see soon.

\paragraph{Backreaction.}
We can estimate the effect of the generated fermions by inserting Eqs.~\eqref{eq:anomaly_induced_result} and \eqref{eq:pair_induced_result} into Eq.~\eqref{eq:eom_w_current}.
The equation of motion including the backreaction can be expressed as
\begin{align}
	0 = \left( Z\,f'' + 2 f^3 \right) - \frac{ 2 \xi_{\text{eff}}}{ ( - \eta )} f^2\,,
	\label{eq:eom_back}
\end{align}
where
\begin{align}
	\xi_{\text{eff}}  = \frac{g^2_{\hat f,\text{eff}} \dot \phi}{8 \pi^2 f_a H}\,,
	\quad
	\frac{1}{g^2_{\hat f,\text{eff}}}
	\equiv \frac{1}{g^2_{\hat f}} + \frac{N_F}{288 \pi^2}\,,
	\quad
	Z \equiv 
	\frac{1+ \frac{\tilde c N_F}{288 \pi^2} g_{\hat f}^2}{1+ \frac{\tilde N_F}{288 \pi^2} g_{\hat f}^2}\,.
\end{align}
Before concluding, let us examine the validity of the key assumption of $f (\eta) \propto 1/ (- \eta)$, under which this expression was derived. We have to check whether the attractor solution of this form still exists in Eq.~\eqref{eq:eom_back}, namely with the backreaction. Interestingly, the equation keeps almost the same form except for $\xi_\text{eff}$ and $Z$. As a result, following the discussion given below Eq.~\eqref{eq:solutions} (see also Ref.~\cite{Domcke:2018rvv}), one may find the condition under which the asymptotic solution of the form $f (\eta) \propto 1 / (-\eta)$ exists and is stable against perturbations. After some manipulation, we find $f (\eta) = c_{2,\text{eff}} \xi_\text{eff} / (- \eta)$ with $c_{2, \text{eff}} = \left(1 + \sqrt{1 - 4 Z / \xi_\text{eff}^2} \right)/2$ under the condition $\xi_\text{eff} > 2 \sqrt{Z}$, which is smoothly connected to the attractor solution without the backreaction \eqref{eq:solutions} for $\xi_\text{eff} \to \xi$ and $Z \to 1$. This a posteriori justifies our self-consistency, namely the use of the specific form of $f(\eta) \propto 1 / (-\eta)$ in Eq.~\eqref{eq:pair_induced_result} as long as $\xi_\text{eff}$ is large enough.

In summary, 
a vacuum contribution to the total induced current
can simply be interpreted as the running of the gauge coupling, evaluated at the background field
value $\hat f$, while a contribution due to chiral fermion excitations tries
to decrease the value of the global minimum for $f(\eta)$ by decreasing $\xi_\text{eff}$. The fermion excitations may further shift the balance between the kinetic and the potential term of $f(\eta)$, altering the relaxation time to reach this global minimum.  However, unless the gauge coupling or the number of fermionic degrees of freedom are very large, the backreaction of the fermion excitations on the background is negligible. This is because we expect $Z \simeq 1$ and $\xi_\text{eff} \simeq \xi$ in this case, indicating that the asymptotic behavior is well approximated by the solution without the backreaction, \textit{i.e.,} $f (\eta) \simeq \xi / (- \eta)$.

\section{Conclusion}
\label{sec:conclusion}

In this paper we study fermion production in a homogeneous and isotropic non-abelian gauge field background. Since such a gauge field background spontaneously breaks $CP$ invariance, we may expect both pair production of fermions (analogous to the Schwinger effect in the presence of a strong electric field) as well as a chiral fermion production channel, resulting in an asymmetric production of left- and right-handed fermions. The latter is directly tied to the chiral anomaly equation, which predicts the generation of chiral charge in gauge field backgrounds with a non-vanishing Chern-Pontryagin density.

Solving the fermion equation of motion, we see how the chiral anomaly equation is explicitly realized from two contributions. Firstly, one mode of the fermion doublet smoothly connects particle and anti-particle states, leading to an asymmetric production of left- and right-handed particles, resulting in the generation of a chiral charge. Secondly, the asymmetry between the dispersion relations of left- and right-handed particles, present in the entire spectrum, results in a non-vanishing vacuum contribution to the chiral charge if the gauge field background is time-dependent. Together these two contributions provide the microphysical explanation for the total chiral charge predicted by the anomaly equation.
{This implies that the naive prescription of normal ordering fails to reproduce the correct result in this case.}

The pair production of particles is on the contrary a non-adiabatic process, sensitive to the details of the  time-evolution of the gauge field background. The homogeneous and isotropic non-abelian gauge field background spontaneously breaks the invariance under global transformations contained in the SU(2) gauge group and the
symmetry of spatial rotations down to a diagonal subgroup. The helicity eigenstates of this subgroup form pairs which mix under the time-evolution of the system. This mixing leads to the pair production of fermions, encoded in non-vanishing Bogolyubov coefficients.

Together, these two fermion production channels fully describe the generation of fermions in a non-abelian gauge field background. The microphysical interpretation provided here allows to extract information which is not contained in the anomaly equation, such as the rate of pair production and the momentum distribution of the generated fermions. This enables us to compute the backreaction of the fermions on the gauge field background, arising through the induced fermion current. Here again, special care is required to correctly account for the vacuum contribution {which in the end amounts to replacing the bare coupling constant with the running gauge coupling.}
We find this backreaction to be small, unless the gauge coupling or the number of fermions is very large. This is a key difference with respect to the abelian counterpart of this study, where the backreaction of the fermions on the gauge field background was found to be very important~\cite{Domcke:2018eki}.

The results presented here shed light on the understanding of the chiral anomaly and fermion pair production in non-abelian gauge theories. This has immediate consequences for settings which predict strong non-abelian gauge fields. One example is chromo-natural inflation, where the (pseudo)scalar field driving cosmic inflation is coupled to the Chern-Pontryagin density. More generally, strong non-abelian gauge fields should be expected in the hot primordial plasma in the early Universe or in the quark-gluon-plasma of heavy ion collisions, and the processes described above will contribute to the thermalization of this plasma. Finally, the chiral charge generated from an adiabatically changing gauge field background may play a role in the generation of the matter antimatter asymmetry of the Universe. We leave the exploration of these exciting questions to future work.


\section*{Acknowledgements}
We thank Kazuya Yonekura for valuable discussions on the eta invariant.
Y.E. is supported in part by a JSPS KAKENHI Grant No.
JP18J00540.

\appendix
\section{Notations and conventions}
\label{sec:nandc}

\paragraph{Metric.}
We take the sign convention $(\eta_{\mu\nu}) = \text{Diag}\, (+,-,-,-)$.
The line element in the Friedmann-Lema\^itre-Robertson-Walker metric with vanishing curvature is
\begin{align}
	\dd s^2 
	= g_{\mu\nu}\, \dd x^\mu \dd x^\nu
	= \dd t^2 - a^2 (t)\, \dd \bm{x}^2 
	= a^2 (\eta) \left( \dd \eta^2 - \dd \bm{x}^2  \right).
\end{align}
We work in conformal coordinates, $(\eta, \bm{x})$, unless otherwise stated.
The vierbein on this background is given by
\begin{align}
	e^\alpha_\mu = a \delta^\alpha_\mu, 
	\quad
	e^\mu_\alpha = \frac{1}{a} \delta^\mu_\alpha.
\end{align}
The totally antisymmetric tensor is defined so that
$\epsilon^{0123} = - \epsilon_{0123} = +1$.

\paragraph{Gamma matrices.}
We adopt the chiral representation for the gamma matrices in Minkowski spacetime:
\begin{align}
	\gamma^0 = \left(\begin{array}{cc}0 & 1 \\1 & 0\end{array}\right),
	\quad
	\bm{\gamma} = \left(\begin{array}{cc}0 & \bm{\sigma} \\ - \bm{\sigma} & 0\end{array}\right),
	\quad
	\gamma_5 = \left(\begin{array}{cc} -1 & 0 \\0 & 1\end{array}\right).
\end{align}
This fulfills the Clifford algebra in Minkowski spacetime
\begin{align}
	\{ \gamma^\mu, \gamma^\nu \} = 2 \eta^{\mu\nu}.
\end{align}
The gamma matrices on the curved background are obtained from
\begin{align}
	\hat \gamma^\mu = e^\mu_\alpha \gamma^\alpha = \frac{\gamma^\mu}{a},
	\quad
	\hat \gamma_\mu = e^\alpha_\mu \gamma_\alpha = a \gamma_\mu.
\end{align}
One may easily see that they satisfy the Clifford algebra on the curved background
\begin{align}
	\{ \hat \gamma^\mu, \hat \gamma^\nu \} = 2 g^{\mu\nu}.
\end{align}

\paragraph{$CP$ transformation.}
We define the $CP$ transformation acting on the Weyl fermions as follows:
\begin{align}
	CP\, \psi_\text{L/R} (x) \, (CP)^{-1} = \mp (i \sigma^2 ) \, T_C \, \psi_\text{L/R}^\dag (x_P),
	\label{eq:cp_weyl}
\end{align}
where $(x_P) = (\eta, - \bm{x})$ {and $\sigma^2$ denotes the second Pauli matrix}. $T_C$ defines a similarity transformation which fulfills $T_C (- T{^a}^\ast) T_C^{-1} = T^a$.
The current which couples to the SU$(2)$ gauge field transforms as
\begin{align}
	CP\, \psi_\text{L/R}^\dag (x) \left( \mp \bm{\sigma} \right) T^a \psi_\text{L/R} (x) \, (CP)^{-1} 
	&= \mp \psi_\text{L/R} (x_P) \left(\sigma^2 \bm{\sigma} \sigma^2 \right) \left( T_C^{-1} T^a T_C \right) \psi_\text{L/R}^\dag (x_P) \nonumber \\
	&= - \psi^\dag_\text{L/R} (x_P) \left( \mp \bm{\sigma} \right) T^a \psi_\text{L/R} (x_P).
\end{align}
To make the gauge interaction invariant under $CP$, we have
\begin{align}
	CP\, \bm{A}^a (x)\, (CP)^{-1} = - \bm{A}^a (x_P).
\end{align}
This indicates the $CP$ transformation of the homogeneous field, $A^a_i = - f \delta^a_i$:
\begin{align}
	CP\, f (\eta) \, (CP)^{-1} = - f (\eta).
	\label{eq:gauge}
\end{align}
The $CP$ transformation acting on the fermion current yields
\begin{align}
	CP\, \left( J_\text{L/R}^0 (x), \bm{J}_\text{L/R} (x) \right) \, (CP)^{-1} 
	&= \psi_\text{L/R} (x_P) \left( \sigma^2 (1, \mp \bm{\sigma}) \sigma^2 \right) \psi_\text{L/R}^\dag (x_P) 
	= \psi_\text{L/R}^\dag (x_P) \left(- 1, \mp \bm{\sigma} \right) \psi_\text{L/R} (x_P)
	\nonumber \\
	&= \left( - J_\text{L/R}^0 (x_P), \bm{J}_\text{L/R} (x_P) \right).
	\label{eq:current_CP}
\end{align}

\section{{Regularization of the fermion current}}
\label{app:regularization}

\subsection{Eta invariant and regularization}
\label{sec:eta_reg}

Here we show explicitly that our regularization for the fermion current does not spoil the $CP$ symmetry.
To avoid unnecessary complications, we discuss the left-handed fermion only.
The application to the right-handed fermion is straightforward.

Let us start with how the $CP$ transformation defined in Eq.~\eqref{eq:cp_weyl} acts on $\psi_\text{L}$, assuming a constant value for $f$.
In this case one may expand the wave function as follows (see Eqs.~\eqref{eq:mode_const_f}):
\begin{align}
	\psi_\text{L} (x) 
	= \int \frac{\dd^3 k}{( 2 \pi )^{3/2}} \, e^{i \bm{k} \cdot \bm{x}}\,
	&\Bigg\{ e^{ - i \omega_\text{L}^{(+1)} \eta} d^{(+1)\, \dag}_{\text{L},- \bm{k}} e_{\bm k}^{(+1)} \nonumber\\[.5em]
	&+  \left[
	e^{-i \omega_\text{L}^{(-1)}  \eta}\, \theta \left(\frac{
	f}{2} - k\right) d^{(-1)\, \dag}_{\text{L},- \bm{k}}
	+  e^{-i \omega_\text{L}^{(-1)}  \eta}\, \theta \left(k - \frac{f}{2} \right) b^{(-1)}_{\text{L},\bm{k}}
	\right]
	e_{\bm k}^{(-1)} \nonumber \\[.5em]
	& + e^{-i \omega_\text{L}^{(0;1)}  \eta} d^{(0)\, \dag}_{\text{L},- \bm{k}} e_{\bm k}^{(0;1)} 
	+ e^{- i \omega_\text{L}^{(0;2)} \eta} b^{(0)}_{\text{L},\bm{k}} e^{(0;2)}_{\bm k} \Bigg\}\,,
	\label{eq:mode_exp_app}
\end{align}
where the energy spectrum is given by
\begin{align}
	\omega_\text{L}^{(+1)} = - \left( k + \frac{f}{2} \right), \quad
	\omega_\text{L}^{(-1)} = k - \frac{f}{2}, \quad
	\omega_\text{L}^{(0;1)} = - \sqrt{k^2 +f^2} + \frac{f}{2}, \quad
	\omega_\text{L}^{(0;2)} = \sqrt{k^2 + f^2} + \frac{f}{2}.
	\nonumber
\end{align}
{ Note that $d_{\text{L}, - \bm k}^{(-1)}$ and $b_{\text{L},\bm k}^{(-1)}$ 
must appear together with the Heaviside theta function since they
are only defined for $k < f / 2$ and $k > f / 2$ respectively.
For a notational simplicity, we usually omit this theta function unless otherwise stated.
First of all, the basis transforms under the $CP$ as follows:
\begin{align}
	&CP\, e_{\bm k}^{(-1)} \, (CP)^{-1} = - i \sigma^2 T_C \, e_{\bm k}^{(+1)\, \ast}\, , \quad
		CP\, e_{\bm k}^{(+1)} \, (CP)^{-1} = - i \sigma^2 T_C \, e_{\bm k}^{(-1)\, \ast}\, ,  \nonumber \\[.5em]
	& CP\, e_{\bm k}^{(0;1)}\, (CP)^{-1} = - i \sigma^2 T_C\,  e_{\bm k}^{(0;2)\, \ast} \, ,
	\quad
	CP\, e_{\bm k}^{(0;2)}\, (CP)^{-1} = - i \sigma^2 T_C\,  e_{\bm k}^{(0;1)\, \ast} \,.
	\label{eq:cp_basis}
\end{align}
To see this, we have used the definition of the basis given in Eq.~\eqref{eq:basis} and the following properties
\begin{align}
	- i \sigma^2 \chi{_{\bm k}^{(\pm)}}^\ast = \chi_{\bm k}^{(\mp)}, \quad
	T_C t{_{\bm k}^{(\pm)}}^\ast = t_{\bm k}^{(\mp)},
\end{align}
which can be shown from $(\hat{\bm{k}} \cdot \bm{\sigma}) \chi_{\bm k}^{(\pm)} = \pm \chi_{\bm k}^{(\pm)}$, $(\hat{\bm{k}} \cdot \bm{T}) t_{\bm k}^{(\pm)} = \pm (1/2) t_{\bm k}^{(\pm)}$, $\sigma^2 \bm{\sigma}^\ast \sigma^2 = - \bm{\sigma}$, and $T_C \bm{T}^\ast T_C^{-1} = - \bm{T}$.
Moreover, using $CP:\, f \mapsto -f$, 
we obtain the $CP$ transformation of the energy levels,
\begin{align}
	 \omega^{(-1)}_\text{L} 
	\overset{CP}\longleftrightarrow 
	- \omega^{(+1)}_\text{L}
	\,, \quad
	\omega^{(0;1)}_\text{L} 	\overset{CP}\longleftrightarrow
	- \omega^{(0;2)}_\text{L}\,. 	\label{eq:cp_energy}
\end{align}
Equipped with Eqs.~\eqref{eq:cp_basis} and \eqref{eq:cp_energy}, we can see the following transformation law by inserting the mode expansion \eqref{eq:mode_exp_app} into the definition of the $CP$ transformation \eqref{eq:cp_weyl}:
\begin{align}
&CP\, \left[ \theta \left( k - \frac{f}{2} \right) b_{\text{L},\bm k}^{(-1)}+ \theta \left( \frac{f}{2} - k \right) d_{\text{L},- \bm k}^{(-1)\, \dag}\right] \, (CP)^{-1} 
	= d_{\text{L},- \bm k}^{(+1)}\,, \nonumber \\[.5em]
	&CP\, d_{\text{L},- \bm k}^{(+1)}\, (CP)^{-1} = \theta \left( k - \frac{f}{2} \right) b_{\text{L},\bm k}^{(-1)} + \theta \left( \frac{f}{2} - k \right) d_{\text{L},- \bm k}^{(-1)\, \dag}, \nonumber \\[.5em]
	&CP\, d_{\text{L},- \bm k}^{(0)}\, (CP)^{-1} = b_{\text{L},\bm k}^{(0)}\,, \quad
	CP\, b_{\text{L}, \bm k}^{(0)}\, (CP)^{-1} = d_{\text{L},- \bm k}^{(0)}\,.
	\label{eq:cp_cran}
\end{align}
Here we have recovered the Heaviside theta function to avoid confusions.
A similar computation yields the following transformation law for the right-handed fermion:
\begin{align}
		&CP\, \left[ \theta\left( k - \frac{f}{2} \right) d_{\text{R}, -\bm k}^{(-1)\, \dag} + \theta \left( \frac{f}{2} - k \right) b_{\text{R},\bm k}^{(-1)} \right]  \, (CP)^{-1} = b_{\text{R},\bm k}^{(+1)\, \dag}\,, \quad \nonumber \\[.5em]
	&CP\, b_{\text{R},\bm k}^{(+1)\, \dag}\, (CP)^{-1} = \theta \left( k - \frac{f}{2} \right) d_{\text{R}, - \bm k}^{(-1)\, \dag} + \theta \left( \frac{f}{2} - k \right) b_{\text{R},\bm k}^{(-1)}\, , \nonumber \\[.5em]
	&CP\, b_{\text{R}, \bm k}^{(0)}\, (CP)^{-1} = d_{\text{R}, -\bm k}^{(0)}\,, \quad
	CP\, d_{\text{R}, - \bm k}^{(0)}\, (CP)^{-1} = b_{\text{R}, \bm k}^{(0)}\,.
\end{align}
}

Now we are in a position to obtain explicitly the CP transformation of the regularized charge [Eq.~\eqref{eq:q_antisym}]:
\begin{align}
	\left[ Q_L \right]_{\hat{\Lambda}}
		= \int \frac{\dd^3 k}{(2 \pi)^3}\, \frac{1}{2} &\Bigg\{ 
	\sum_{p} R \left(\frac{|\omega_L^{(p)}|}{{a\hat{\Lambda}}} \right)
	\left[ b^{(p)\, \dag}_{\text{L}, \bm k} b{^{(p)}_{\text{L},\bm k}} 
	- b{^{(p)}_{\text{L},\bm k}} b^{(p)\, \dag}_{\text{L}, \bm k} 
	\right] \nonumber\\
	& \qquad -
	\sum_{n} R \left(\frac{|\omega_L^{(n)}|}{{a\hat{\Lambda}}} \right)
	\left[ d^{(n)\, \dag}_{\text{L},- \bm k} d{^{(n)}_{\text{L}, - \bm k}} 
	- d^{(n)}_{\text{L},- \bm k} d^{(n)\, \dag}_{\text{L},- \bm k} 
	\right]
	\Bigg\}
	 \nonumber \\[.5 em]
	 = \int \frac{\dd^3 k}{(2 \pi)^3}\, \Bigg\{ &
	 	- R \left( \frac{| \omega^{(+1)}_\text{L} |}{{a\hat{\Lambda}}} \right)
		\left[ d_{\text{L},- \bm k}^{(+1)\, \dag} d_{\text{L}- \bm k}^{(+1)} - \frac{1}{2} \text{vol}\, (\mathbb{R}^3) \right] \label{eq:+1} \\
		& + R \left( \frac{| \omega^{(-1)}_\text{L} |}{{a\hat{\Lambda}}} \right)
		\left[ b_{\text{L},\bm k}^{(-1)\, \dag} b_{\text{L},\bm k}^{(-1)} - d_{\text{L}, - \bm k}^{(-1)\, \dag} d_{\text{L}, - \bm k}^{(-1)} - \text{sgn}\, \left( \omega_\text{L}^{(-1)} \right) \frac{1}{2} \text{vol}\, (\mathbb{R}^3) \right] \label{eq:-1} \\
		& - R \left( \frac{| \omega^{(0;1)}_\text{L} |}{{a\hat{\Lambda}}} \right)
		\left[ d_{\text{L},- \bm k}^{(0)\, \dag} d_{\text{L}- \bm k}^{(0)} - \frac{1}{2} \text{vol}\, (\mathbb{R}^3) \right] \label{eq:01} \\
		&+ R \left( \frac{| \omega^{(0;2)}_\text{L} |}{{a\hat{\Lambda}}} \right)
		\left[ b_{\text{L},\bm k}^{(0)\, \dag} b_{\text{L},\bm k}^{(0)} - \frac{1}{2} \text{vol}\, (\mathbb{R}^3) \right] 
		\Bigg\}
		\label{eq:02}
\end{align}
Here $(p)$ runs over $(-1)$ and $(0;2)$ while $(n)$ does $(+1)$, $(-1)$, and $(0;1)$ (see Fig.~\ref{fig:dispersion}).
By using Eqs.~\eqref{eq:cp_energy} and \eqref{eq:cp_cran}, one can easily confirm $\eqref{eq:+1} \overset{CP}\leftrightarrow - \eqref{eq:-1}$ and $\eqref{eq:01} \overset{CP} \leftrightarrow - \eqref{eq:02}$, which means
\begin{align}
	CP\, \left[ Q_\text{L} \right]_{\hat{\Lambda}} \, (CP)^{-1}
	= - \left[ Q_\text{L} \right]_{\hat{\Lambda}}.
\end{align}

\subsection{Independence of regularization}
\label{sec:indep}

Here we show that the result given in Eq.~\eqref{eq:Qv_const} does not depend on the {explicit functional form} of the regularization.
We assume that $R(0) = 1$, {the $n$-th derivative} $R^{(n)}$ is regular at $x = 0$, and $R^{(n)}$ drops rapidly for $x \to \infty$, namely faster than a polynomial of $x$.
Let us start with the definition of the eta invariant:
\begin{align}
		Q_H^\text{(v)} \equiv \lim_{{\hat{\Lambda}} \to \infty}
		[ Q_H^\text{(v)} ]_{\hat \Lambda}
		\equiv \lim_{{\hat{\Lambda}} \to \infty} \text{vol}\, (\mathbb{R}^3) \int \frac{\dd^3 k}{(2\pi)^3}\,
	\left[ 
		- \frac{1}{2} \sum_\lambda \text{sgn}\, \left( \omega^{(\lambda)}_{H} \right) \,  R \left( \frac{ | \omega^{(\lambda)}_H |}{{a \hat{\Lambda}}}\right)
	\right] \,.
\end{align}
By inserting Eqs.~\eqref{eq:disp_++}--\eqref{eq:disp_02},
one finds
\begin{align}
	[ Q_\text{L/R}^\text{(v)} ]_{\hat \Lambda}
	=
	\mp \text{vol}\, (\mathbb{R}^3)\, \frac{f^3}{4\pi^2}\, 
	\Bigg\{
	- \frac{1}{12}  + 
	\frac{1}{\epsilon^3} \int \dd x\, x^2
	\Bigg[ &
		R \left( x - \frac{\epsilon}{2} \right)
		- R \left( x + \frac{\epsilon}{2} \right) \nonumber \\[.5em]
		&+ R \left( \sqrt{x^2 + \epsilon^2} + \frac{\epsilon}{2}\right)
		- R \left( \sqrt{x^2 + \epsilon^2} - \frac{\epsilon}{2} \right)
	 \Bigg]
	 \Bigg\}\,,
\end{align}
where $\epsilon = f / (a \hat \Lambda) \ll 1$ and $x = \epsilon k / f$.
This integral gives a meaningful result only after the regularization.

To see how $R$ regulates this integral, we separate the integral into two parts:
(i) $0 \leq x \leq M \epsilon \ll1$ and (ii) $M \epsilon \leq x$ with $M \gg 1$.
One may take $M$ to be arbitrary large while keeping $M \epsilon \ll 1$ for a sufficiently small $\epsilon$.
Thus, in the regime (ii), one may expand the integral by $\epsilon / x \leq 1/ M \ll 1$ and just take the leading part, which gives
\begin{align}
	\frac{1}{\epsilon^3}\int_{M \epsilon} \dd x\, R'' (x)\, \frac{\epsilon^3 x}{2}
	= \Big[ R' (x) \frac{x}{2} \Big]^\infty_{M \epsilon} - 
	\Big[ \frac{R(x)}{2} \Big]^\infty_{M \epsilon}
	\; \overset{\epsilon \to 0} \longrightarrow \; \frac{1}{2} \,.
\end{align}
Here we have used the following property of $R$:
$x R'(x) \to 0$ for $x\to 0$; $ x R'(x) \to 0$, $R(x) \to 0$ for $x \to \infty$; and $R(0) = 1$.
In the regime (i), one may expand the integrand by both $\epsilon$ and $x$ because of $x \leq M \epsilon \ll 1$ and $\epsilon \ll 1$.
Then one obtains
\begin{align}
	\frac{1}{\epsilon^3} \int_0^{M \epsilon} \dd x\, x^2\,
	\left[
		\epsilon R''(0) \left( \sqrt{x^2 + \epsilon^2} - x \right)
	\right] + \mathcal{O} (\epsilon^2)
	= \frac{M^3 \epsilon}{3} R'' (0) + \mathcal{O} (\epsilon^2) \;
	\overset{\epsilon \to 0}\longrightarrow \; 0\,.
\end{align}

Therefore we get the following result without specifying the regulator $R$:
\begin{align}
	Q_\text{L/R}^\text{(v)} = 
	\frac{5}{6}\times \left[\mp \text{vol}\, (\mathbb{R}^3)\, \frac{f^3}{8\pi^2} \right]\,.
\end{align}

\section{{General representations of fermions}}
\label{sec:general}
{Here we summarize results for fermions in a general spin-$j$ representation of the SU(2) gauge group. The gauge field background~\eqref{eq:gauge_bkg_hom} spontaneously breaks the SU(2) gauge group and the SO(3) group of spatial rotations down to the diagonal SO(3) subgroup. With respect to the axis specified by the momentum vector $\vec k$ (taken here w.l.o.g.\ to be the $z$-direction), this symmetry is further reduced to the SO(2) symmetry associated with helicity. The corresponding conserved charge is the projection of the total spin onto the $z$-axis, $s/2 + m$, with $s/2 = \pm 1/2$ denoting the intrinsic spin of the fermion under spatial rotations and $m = -j, -(j - 1), \dots j$ accounting for the $z$-component of the spin $j$ associated with the $SU(2)$ gauge symmetry.
The results for the fermion doublet (fundamental representation) discussed in the main text are obtained by setting $j = 1/2$.\footnote{For notational ease, we use $m \mapsto 2 m $ in the main text.} The generalization to higher representations presented here helps to clarify the conceptual structure of the computations in the main text.}

\paragraph{Equation of motion.}
The fermion equation of motion is given by Eq.~\eqref{eq:eom_fermion0}:
\begin{align}
	0 = \left[ i \partial_\eta \pm \bm{\sigma} \cdot \bm{k} \pm f(\eta)\, 
	\bm{\sigma} \cdot \bm{T} \right] \psi_\text{L/R} (\eta, \bm{k})\,.
\end{align}
For a spin-$j$ representation, the eigenbases of the spin and gauge degrees of freedom are given by
\begin{align}
	\left(\hat{\bm{k}}\cdot\bm{\sigma}\right)\chi_{\bm{k}}^{(\pm)} = \pm \chi_{\bm{k}}^{(\pm)}\,,
	~~~
	\left(\hat{\bm{k}}\cdot\bm{T}\right)t_{\bm{k}}^{(m)} = m t_{\bm{k}}^{(m)}\,,
\end{align}
where $m$ runs over $-j, -j+1, ..., j$. The wave function is expanded as
\begin{align}
	\psi_\text{L/R}(\eta, \bm{k}) = 
	\sum_{s=\pm, m} \psi_\text{L/R}^{(s,m)}(\eta, \bm{k})\chi_{\bm{k}}^{(s)}t_{\bm{k}}^{(m)}\,.
\end{align}
The equation of motion for each mode is given by 
\begin{align}
	0 &= \left[i \partial_\eta \pm \left(k + j f(\eta)\right) \right]\psi_{\text{L/R}}^{(+, j)}(\eta, \bm{k})\,, \\
	0 &= \left[i \partial_\eta \mp \left(k - j f(\eta)\right) \right]\psi_{\text{L/R}}^{(-, -j)}(\eta, \bm{k})\,, \\
	0 &= \left[i \partial_\eta \pm k\begin{pmatrix} 1 & \\ & -1 \end{pmatrix} 
	\pm f(\eta) \begin{pmatrix} m-1 & \sqrt{(j+1/2)^2 - (m-1/2)^2} \\ 
	\sqrt{(j+1/2)^2 - (m-1/2)^2} & -m \end{pmatrix} \right]
	\begin{pmatrix} \psi_{\text{L/R}}^{(+, m-1)} \\ \psi_{\text{L/R}}^{(-, m)} \end{pmatrix}\,,
\end{align}
where $m$ now runs $-j+1, -j+2, ..., j$. 
Note that the diagonal part of the SU(2) and SO(3) remains as a symmetry, 
resulting in the conservation of $s/2 + m$. 
Thus we can understand the above structure 
by realizing that only the modes with the same value of $s/2 + m$ mix with each other.

\paragraph{Energy eigenbasis for constant $f$.}
For constant $f$, the wave functions satisfy the plane wave solution with the dispersion relations given by
\begin{align}
	\omega_{\text{L/R}}^{(+\tilde{\jmath})} &= \mp (k + jf)\,, \\
	\omega_{\text{L/R}}^{(-\tilde{\jmath})} &= \pm (k - jf)\,, \\
	\omega_{\text{L/R}}^{(\tilm; 1)} &= 
	\mp \left(\sqrt{\left(k+\tilm f\right)^2 + f^2 \left(\tilde{\jmath}^2 - \tilm^2\right)} - \frac{f}{2}\right)\,, \\
	\omega_{\text{L/R}}^{(\tilm; 2)} &= 
	\pm \left(\sqrt{\left(k+\tilm f\right)^2 + f^2 \left(\tilde{\jmath}^2 - \tilm^2\right)} + \frac{f}{2}\right)\,,
\end{align}
for $\tilm = -\tilde{\jmath} + 1, -\tilde{\jmath} + 2, ..., \tilde{\jmath} - 1$, 
where we have defined $\tilm \equiv m -1/2$ and $\tilde{\jmath} \equiv j + 1/2$ which refer to the total spins.
The orthogonal matrix that diagonalizes the $\tilm$ modes is given by
\begin{align}
	O_{\tilm}(\kappa) = 
	\begin{pmatrix} 
	\cos\theta_{\tilm} & \sin \theta_{\tilm} \\
	-\sin \theta_{\tilm} & \cos\theta_{\tilm} 
	\end{pmatrix}\,,
	\quad
	\tan \left[2\theta_{\tilm}(\kappa)\right] = \frac{\sqrt{\tilde{\jmath}^2 - \tilm^2}}{\kappa + \tilm}\,,
\end{align}
so that
\begin{align}
	\begin{pmatrix}
	\psi_{\text{L/R}}^{(\tilm; 1)}(\eta, \bm{k}) \\
	\psi_{\text{L/R}}^{(\tilm; 2)}(\eta, \bm{k})
	\end{pmatrix}
	=
	O_{\tilm}\left(k/f\right)
	\begin{pmatrix}
	\psi_{\text{L/R}}^{(+, m-1)}(\eta, \bm{k}) \\
	\psi_{\text{L/R}}^{(-, m)}(\eta, \bm{k})
	\end{pmatrix}\,.
\end{align}
The wave function for the left-handed fermion expanded in terms of creation and annihilation operators is given by
\begin{align}
	\psi_{\text{L}} &= e^{-i\omega_{\text{L}}^{(+\tilj)}\eta}{{d^{(+\tilj)}_{\text{L},-\bm{k}}}}^\dagger e_{\bm{k}}^{(+\tilj)}
	+ e^{-i\omega_{\text{L}}^{(-\tilj)}\eta}
	\left[ \theta\left(k - jf\right) b_{\text{L}, \bm{k}}^{(-\tilj)} + \theta\left(jf - k\right) 
	{d_{\text{L},-\bm{k}}^{(-\tilj)}}^\dagger
	\right]e_{\bm{k}}^{(-\tilj)} \nonumber \\
	&+ \sum_{\tilm = -\tilj +1}^{\tilj - 1}
	\left[
	e^{-i\omega_{\text{L}}^{(\tilm; 1)}\eta} 
	{d_{\text{L},-\bm{k}}^{(\tilm)}}^\dagger e_{\bm{k}}^{(\tilm; 1)}
	+ e^{-i\omega_{\text{L}}^{(\tilm; 2)}\eta}
	b_{\text{L},\bm{k}}^{(\tilm)} e_{\bm{k}}^{(\tilm; 2)}
	\right]\,,
\end{align}
where the energy eigenbasis is defined as
\begin{align}
	e_{\bm{k}}^{(+\tilj)} \equiv \chi_{\bm{k}}^{(+)}t_{\bm{k}}^{(+j)}\,,
	\quad
	e_{\bm{k}}^{(-\tilj)} \equiv \chi_{\bm{k}}^{(-)}t_{\bm{k}}^{(-j)}\,,
	\quad
	\begin{pmatrix}
	e_{\bm{k}}^{(\tilm; 1)} \\
	e_{\bm{k}}^{(\tilm; 2)}
	\end{pmatrix}
	=
	O_{\tilm}\left(k/f\right)
	\begin{pmatrix}
	\chi_{\bm{k}}^{(+)}t_{\bm{k}}^{(m-1)} \\
	\chi_{\bm{k}}^{(-)}t_{\bm{k}}^{(m)}
	\end{pmatrix}\,,
\end{align}
for $\tilm = -\tilj+1, ..., \tilj -1$.
A similar expansion holds for the right-handed fermion.

\paragraph{$CP$ transformation.}
The basis transforms under the $CP$ as
\begin{align}
	&CP\, e_{\bm k}^{(-\tilj)} \, (CP)^{-1} = - i \sigma^2 T_C \, e_{\bm k}^{(+\tilj)\, \ast}\, , \quad
		CP\, e_{\bm k}^{(+\tilj)} \, (CP)^{-1} = - i \sigma^2 T_C \, e_{\bm k}^{(-\tilj)\, \ast}\, ,  \nonumber \\[.5em]
	& CP\, e_{\bm k}^{(\tilm;1)}\, (CP)^{-1} = - i \sigma^2 T_C\,  e_{\bm k}^{(-\tilm;2)\, \ast} \, ,
	\quad
	CP\, e_{\bm k}^{(\tilm;2)}\, (CP)^{-1} = - i \sigma^2 T_C\,  e_{\bm k}^{(-\tilm;1)\, \ast} \,,
\end{align}
where we have used the fact that $O_{\tilm}^\dagger(\kappa) = O_{-\tilm}(-\kappa)$.
The frequency is transformed as
\begin{align}
	\omega_{{H}}^{(-\tilj)}
	\overset{CP}\longleftrightarrow
	-\omega_{{H}}^{(+\tilj)}\,,
	\quad
	\omega_{{H}}^{(\tilm; 1)}
	\overset{CP}\longleftrightarrow
	-\omega_{{H}}^{(-\tilm; 2)}\,.
\end{align}
The $CP$ transformation of the creation and annihilation operators is accordingly given by
\begin{align}
&CP\, \left[ \theta \left( k - j f\right) b_{\text{L},\bm k}^{(-\tilj)}+ \theta \left(j f - k \right) d_{\text{L},- \bm k}^{(-\tilj)\, \dag}\right] \, (CP)^{-1} 
	= d_{\text{L},- \bm k}^{(+\tilj)}\,, \nonumber \\[.5em]
	&CP\, d_{\text{L},- \bm k}^{(+\tilj)}\, (CP)^{-1} = \theta \left( k - j f \right) b_{\text{L},\bm k}^{(-\tilj)} + \theta \left( j f - k \right) d_{\text{L},- \bm k}^{(-\tilj)\, \dag}, \nonumber \\[.5em]
	&CP\, d_{\text{L},- \bm k}^{(\tilm)}\, (CP)^{-1} = b_{\text{L},\bm k}^{(-\tilm)}\,, \quad
	CP\, b_{\text{L}, \bm k}^{(\tilm)}\, (CP)^{-1} = d_{\text{L},- \bm k}^{(-\tilm)}\,,
	\label{eq:cp_cran_gen}
\end{align}
and
\begin{align}
		&CP\, \left[ \theta\left( k - j f \right) d_{\text{R}, -\bm k}^{(-\tilj)\, \dag} + \theta \left( j f - k \right) b_{\text{R},\bm k}^{(-\tilj)} \right]  \, (CP)^{-1} = b_{\text{R},\bm k}^{(+\tilj)\, \dag}\,, \quad \nonumber \\[.5em]
	&CP\, b_{\text{R},\bm k}^{(+\tilj)\, \dag}\, (CP)^{-1} = \theta \left( k - j f \right) d_{\text{R}, - \bm k}^{(-\tilj)\, \dag} + \theta \left( j f - k \right) b_{\text{R},\bm k}^{(-\tilj)}\, , \nonumber \\[.5em]
	&CP\, b_{\text{R}, \bm k}^{(\tilm)}\, (CP)^{-1} = d_{\text{R}, -\bm k}^{(-\tilm)}\,, \quad
	CP\, d_{\text{R}, - \bm k}^{(\tilm)}\, (CP)^{-1} = b_{\text{R}, \bm k}^{(-\tilm)}\,,
\end{align}
where we have written the Heaviside theta explicitly to avoid confusions.
The current is antisymmetrized in the same way as the main text to respect the $CP$ transformation:
\begin{align}
	\left[ Q_H \right]_{\hat{\Lambda}}
	= \int \frac{\dd^3 k}{(2 \pi)^3}\, \frac{1}{2}
	\Bigg\{ &
	\sum_{p} R \left(\frac{|\omega_H^{(p)}|}{{a \hat{\Lambda}}} \right)
	\left[ b^{(p)\, \dag}_{H,\bm k} b{^{(p)}_{H,\bm k}} 
	- b{^{(p)}_{H,\bm k}} b^{(p)\, \dag}_{H,\bm k}
	\right] \nonumber \\
	& \qquad  -
	\sum_{n} R \left(\frac{|\omega_H^{(n)}|}{a{\hat{\Lambda}}} \right)
	\left[ d^{(n)\, \dag}_{H,- \bm k} d{^{(n)}_{H,- \bm k}} - 
	d{^{(n)}_{H,- \bm k}} d^{(n)\, \dag}_{H,- \bm k}
	\right]
	\Bigg\} \,.
\end{align}
We can divide it into the contributions from the excitation and the vacuum in the same way as before,
resulting in
\begin{align}
	Q_H 
	= \lim_{{\hat{\Lambda}} \to \infty} \left[ Q_H \right]_{\hat{\Lambda}}
	=\normord{Q_H}
	+ \lim_{{\hat{\Lambda}} \to \infty} \text{vol}\, (\mathbb{R}^3) \int \dd^3 k\,
	\left[ 
	- \frac{1}{2} \sum_\lambda \text{sgn}\, \left( \omega^{(\lambda)}_{H} \right) 
	\,  R \left( \frac{|\omega^{(\lambda)}_H|}{{a \hat{\Lambda}}}\right)
	\right] \,.
\end{align}

\paragraph{Chiral asymmetry.}
The computation of the chiral asymmetry proceeds completely analogous to Sec.~\ref{sec:anomaly}. 
The contribution from the excitation is given by
\begin{align}
	\delta Q_{\text{L/R}}^\text{(e)} 
	=
	\mp \text{vol}\, (\mathbb{R}^3)\, \frac{j^3}{6\pi^2}
	 \left( f_f^3 - f_i^3 \right)\,,
\end{align}
while the contribution from the vacuum is given by
\begin{align}
	\delta Q_{\text{L/R}}^\text{(v)}
	=
	\mp \frac{\text{vol}\, (\mathbb{R}^3)}{4\pi^2} 
	\left(T(\textbf{2j+1}) - \frac{2}{3}j^3\right)
	\left( f_f^3 - f_i^3\right)\,,
\end{align}
such that
\begin{align}
	\delta Q_{\text{L/R}}^\text{(e)} + \delta Q_{\text{L/R}}^\text{(v)}
	=
	\mp \text{vol}\,(\mathbb{R}^3)\, \frac{T(\textbf{2j+1})}{4\pi^2}
	\left(f_f^3 - f_i^3\right)\,.
\end{align}
Thus it correctly reproduces the anomaly equation for a general spin-$j$ representation.

\paragraph{Pair production in CNI background.}
The production in the modes with $\tilm = \pm \tilj$ is not affected by the non-adiabatic evolution $f' \neq 0$.
Thus we concentrate on the modes with $\tilm = -\tilj + 1, ..., \tilj - 1$, \textit{i.e.}, the mixed modes.
In the following we assume, as in Sec.~\ref{sec:app_cni}, that $f = \xi/(-\eta)$.
Then the equation of motion in the energy eigenbasis for constant $f$ is given by
\begin{align}
	0 &= 
	\left[i\partial_\eta - \begin{pmatrix} \omega_{\text{L/R}}^{(\tilm; 1)} & \\ & \omega_{\text{L/R}}^{(\tilm; 2)} \end{pmatrix}
	+ i\left(\partial_\eta \theta_{\tilm}\right) \begin{pmatrix} & -1 \\ 1 & \end{pmatrix}
	\right]
	\begin{pmatrix}
	\psi_{\text{L/R}}^{(\tilm; 1)}(\eta, \bm{k}) \\
	\psi_{\text{L/R}}^{(\tilm; 2)}(\eta, \bm{k})
	\end{pmatrix}
	\,, \\
	\partial_\eta \theta_{\tilm} &= 
	\frac{k}{2f} \frac{\sqrt{\tilj^2 - \tilm^2}}{(k/f + \tilm)^2 + \tilj^2 - \tilm^2} \frac{f'}{f}\,.
\end{align}
Once we define
\begin{align}
	\begin{pmatrix}
	\psi_{\text{L/R}}^{(\tilm; 1)} \\
	\psi_{\text{L/R}}^{(\tilm; 2)}
	\end{pmatrix}
	\equiv
	\begin{pmatrix}
	e^{-i\int^\eta \omega_{\text{L/R}}^{(\tilm; 1)}}\varphi_{\text{L/R}}^{(\tilm; 1)} \\
	e^{-i\int^\eta \omega_{\text{L/R}}^{(\tilm; 2)}}\varphi_{\text{L/R}}^{(\tilm; 2)} \\
	\end{pmatrix}\,,
\end{align}
they satisfy
\begin{align}
	\partial_\eta \varphi_{\text{L/R}}^{(\tilm; 1)}
	&= \left(\partial_\eta \theta_{\tilm}\right)e^{\mp 2i \Theta_{\tilm}}\varphi_{\text{L/R}}^{(\tilm; 2)}\,,
	\quad
	\partial_\eta \varphi_{\text{L/R}}^{(\tilm; 2)}
	= -\left(\partial_\eta \theta_{\tilm}\right)e^{\pm 2i \Theta_{\tilm}}\varphi_{\text{L/R}}^{(\tilm; 1)}\,,
	\label{eq:bogo_eom}
\end{align}
where $\Theta_{\tilm} = \int^{\eta}d\eta \sqrt{(k+\tilm f)^2 + f^2(\tilj^2 - \tilm^2)}$\,.
From now we concentrate on the left-handed fermion for concreteness.
The Bogolyubov transformation is given by
\begin{align}
	B_{\text{L},\bm{k}}^{(\tilm)} = \alpha_{\text{L},\bm{k}}^{(\tilm)}b_{\text{L},\bm{k}}^{(\tilm)}
	-{\beta_{\text{L},\bm{k}}^{(\tilm)}}^* {d_{\text{L},\bm{-k}}^{(\tilm)}}^\dagger \,,
	\quad
	{D_{\text{L},\bm{k}}^{(\tilm)}}^\dagger = \beta_{\text{L},\bm{k}}^{(\tilm)}b_{\text{L},\bm{k}}^{(\tilm)}
	+{\alpha_{\text{L},\bm{k}}^{(\tilm)}}^*{d_{\text{L},\bm{-k}}^{(\tilm)}}^\dagger \,,
\end{align}
where the coefficients are defined as
\begin{align}
	\alpha_{\text{L},\bm{k}}\left(\eta_f\right) \equiv \varphi_{\text{L},+}^{(\tilm; 2)}\left(\eta_f, \bm{k}\right)\,,
	\quad
	\beta_{\text{L},\bm{k}}\left(\eta_f\right) \equiv \varphi_{\text{L,}+}^{(\tilm; 1)}\left(\eta_f, \bm{k}\right)\,,
\end{align}
with $\varphi_{\text{L},+}^{(\tilm; 1)} = 0$ and $\varphi_{\text{L},+}^{(\tilm; 2)} = 1$ at $\eta = \eta_i$.
The Bogolyubov coefficient is estimated as
\begin{align}
	\abs{\beta_{H}^{(\tilm)}} \sim 
	\begin{cases}
	\displaystyle \frac{k(-\eta_i)}{2 \tilj \xi}\sqrt{\frac{1 - \tilm^2/\tilj^2}{1+4\tilj^2\xi^2}}
	\quad&\mathrm{for}\quad
	k \ll \displaystyle \frac{\tilj \xi}{-\eta_i}\,, \\[1em]
	{\displaystyle \frac{c}{4\tilj \xi} \sqrt{\frac{1 - \tilm^2/\tilj^2}{(1+\tilm/\tilj)^3}}}
	&\mathrm{for}\quad
	\displaystyle \frac{\tilj \xi }{-\eta_i} \ll k \ll \displaystyle \frac{\tilj \xi}{-\eta_f}\,, 
	\label{eq:beta_peak_general} \\[1em]
	\displaystyle \frac{\tilj \xi}{4 (k\eta_f)^2} \sqrt{1 - \frac{\tilm^2}{\tilj^2}}
	&\mathrm{for}\quad
	k \gg \displaystyle \frac{\tilj \xi}{-\eta_f}\,,
	\end{cases}
\end{align}
{with $c \sim 0.4$. 
The behavior for the large and small momentum limit is understood analytically (see App.~\ref{sec:bogogo}),
while the middle one is estimated by numerical computation.}
Compared to the production via the chiral anomaly, the number density is suppressed by $1/\tilj^2 \xi^2$.

\paragraph{Induced current.}
The regularized left-handed induced current for a general spin-$j$ representation is given by
\begin{align}
	\left[ \K_\text{L} \right]_{\hat{\Lambda}}
	\equiv 
	\int \frac{\dd^3 k}{( 2 \pi)^3}\, \Bigg\{ &
		- R \left( \frac{|\omega^{(+\tilj)}_\text{L}|}{{a\hat{\Lambda}}} \right) 
		j \left[ D_{\text{L},- \bm k}^{(+\tilj)\, \dag} D_{\text{L},- \bm k}^{(+\tilj)} 
		- \frac{1}{2} \text{vol}\, (\mathbb{R}^3) \right] 
		\nonumber \\[.5em]
		& + R \left( \frac{| \omega^{(-\tilj)}_\text{L} |}{{a\hat{\Lambda}}} \right) 
		j \left[ B_{\text{L}, \bm k}^{(-\tilj)\, \dag} B_{\text{L}, \bm k}^{(-\tilj)} 
		- D_{\text{L},- \bm k}^{(-\tilj)\, \dag} D_{\text{L},- \bm k}^{(-\tilj)} 
		- \text{sgn}\, \left(\omega^{(-\tilj)}_\text{L} \right)\, \frac{1}{2} \text{vol}\, (\mathbb{R}^3) \right] 
		\nonumber \\[.5em]
		+\sum_{\tilm = -\tilj +1}^{\tilj -1}
		 &\left[ - R \left( \frac{|\omega^{(\tilm; 1)}_\text{L}|}{{a\hat{\Lambda}}} \right) 
		 \left( \frac{\tilj^2 + \tilm(k/f)}{\sqrt{(k/f + \tilm)^2 + \tilj^2 - \tilm^2}} - \frac{1}{2} \right)
		 \left[D^{(\tilm)\,\dag}_{\text{L}, - \bm k} D^{(\tilm)}_{\text{L},- \bm k} 
		 - \frac{1}{2} \text{vol}\, (\mathbb{R}^3)
		 \right] \right.
		 \nonumber\\[.5em]
		 &\left.+ R \left( \frac{|\omega^{(\tilm; 2)}_\text{L}|}{{a\hat{\Lambda}}} \right) 
		\left( -\frac{\tilj^2 + \tilm(k/f)}{\sqrt{(k/f + \tilm)^2 + \tilj^2- \tilm^2}} - \frac{1}{2} \right)
		 \left[
		 	B^{(\tilm)\,\dag}_{\text{L}, \bm k} B^{(\tilm)}_{\text{L},\bm k} - \frac{1}{2} \text{vol}\, (\mathbb{R}^3)
		 \right] 
		 \right.\nonumber\\[.5em]
		 &\left.+ \frac{k}{2f}\sqrt{\frac{\tilj^2 - \tilm^2}{(k/f + \tilm)^2 + \tilj^2 - \tilm^2}} 
		 \left[ R \left( \frac{|\omega^{(\tilm; 1)}_\text{L}|}{{a\hat{\Lambda}}} \right) 
		 + R \left( \frac{|\omega^{(\tilm; 2)}_\text{L}|}{{a\hat{\Lambda}}} \right) \right] 
		 \left( e^{2 i \Theta_{\tilm}} B^{(\tilm)\,\dag}_{\text{L},\bm k} D^{(\tilm)\, \dag}_{\text{L}, - \bm k} 
		 + \text{H.c.} \right)
	\right]
	\Bigg\} \,.
\end{align}
The Bogolyubov coefficient relevant for the computation of the divergent part is given by
\begin{align}
	\alpha_{\text{L},\bm{k}} \simeq 1\,,
	\quad
	\beta_{\text{L},\bm{k}} \simeq 
	\frac{\tilj \xi e^{-2ik\eta}}{4}\sqrt{1 - \frac{\tilm^2}{\tilj^2}}\left(\frac{i}{k^2\eta^2} - \frac{1}{k^3\eta^3} \right)\,,
	\quad
	\Theta_{\tilm} \simeq k\eta\,.
\end{align}
Then we can compute the divergent part of the vacuum contribution in exactly the same way as the main text:
\begin{align}
	[ \K_\text{L}^\text{(v)} ]_{\hat{\Lambda}} \simeq
	\text{vol}\, (\mathbb{R}^3) \times 
	\frac{T(\textbf{2j+1})}{8 \pi^2} \ln \left( \frac{\hat{f}^2}{{\hat{\Lambda}}^2} \right)\, \left( f'' + 2 f^3 \right)\,.
\end{align}
It again describes the running of the gauge coupling from the fermion loop.
The induced current from the excitations is given by
\begin{align}
	\vev{\normord{\K_\text{L}}} = 
	\int \frac{\dd^3 k}{(2 \pi )^3} \, \Bigg\{ &
		j \vev{ B^{(+\tilj)\,\dag}_{\text{L}, \bm k} B^{(+\tilj)}_{\text{L}, \bm k} 
		+ B^{(-\tilj)\,\dag}_{\text{L}, \bm k} B^{(-\tilj)}_{\text{L}, \bm k} 
		- D^{(-\tilj)\,\dag}_{\text{L}, \bm k} D^{(-\tilj)}_{\text{L}, \bm k} } \nonumber \\[.5em]
	\sum_{\tilm = -\tilj + 1}^{\tilj -1} & \left[
	- \left( \frac{\tilj^2 + \tilm (k/f)}{\sqrt{(k/f+\tilm)^2 + \tilj^2 - \tilm^2}} - \frac{1}{2} \right)
		 	\vev{D^{(\tilm)\,\dag}_{\text{L}, \bm k} D^{(\tilm)}_{\text{L}, \bm k}}
	\right. \nonumber \\[.5em] 
	&~~ \left. + \left( -\frac{\tilj^2 + \tilm (k/f)}{\sqrt{(k/f+\tilm)^2 + \tilj^2 - \tilm^2}} - \frac{1}{2} \right)
	\vev{B^{(\tilm)\,\dag}_{\text{L}, \bm k} B^{(\tilm)}_{\text{L}, \bm k}}
	\right]
	\Bigg\}\,.
\end{align}
There are two contributions as before. The contribution from the anomaly is 
\begin{align}
	\vev{ \normord{\K_\text{L}} }_\text{anomaly}
	= - \text{vol}\, (\mathbb{R}^3) \,
	\frac{j^4}{6 \pi^2} f^3\,.
	\label{eq:KLaj}
\end{align}
The contribution from the pair production is
\begin{align}
	\vev{ \normord{\K_\text{L}} }_\text{pair} \sim - \text{vol}\, (\mathbb{R}^3)\, \frac{\tilde{c} j^3}{12 \pi^2} f''\,,
	\label{eq:KLpj}
\end{align}
{where we numerically find that $\tilde{c}\sim \mathcal{O}(0.1)$
almost independent of $j$.}\footnote{
{The coefficient $\tilde{c}$ tends to change more for lower $j$, indicating that there are also terms proportional to $j^2$, $j$, and so on.
}}
They may affect the dynamics of CNI for large enough $j$, 
but the gauge coupling may blow up soon for such a large representation.

\section{Analytic estimation of Bogolyubov coefficients}
\label{sec:bogogo}
Here we derive an analytic estimation of the Bogolyubov coefficients in the small and large momentum limit.
The latter is necessary to evaluate the structure of the divergence in our theory.
We consider a general spin-$j$ representation in this appendix. The results for the doublet fermions used 
in the main text are obtained by setting $\tilj = 1$ and $\tilm = 0$, where $\tilj \equiv j + 1/2$ and $\tilm \equiv m - 1/2$
with $m$ being the $z$-component of the spin associated with the SU(2) gauge symmetry.
We will focus on the left-handed fermion just for notational ease,
but the same result is obtained for the right-handed fermion.

The time evolution of the Bogolyubov coefficients is governed 
by Eq.~\eqref{eq:bogo_eom}, or
\begin{align}
	\partial_\eta \beta_{\text{L},\bm{k}}^{(\tilm)} 
	&= \left(\partial_\eta \theta_{\tilm} \right) e^{-2i\Theta_{\tilm}} \alpha_{\text{L},\bm{k}}^{(\tilm)}\,,
	\quad
	\partial_\eta \alpha_{\text{L},\bm{k}}^{(\tilm)} 
	= -\left(\partial_\eta \theta_{\tilm} \right) e^{+2i\Theta_{\tilm}} \beta_{\text{L},\bm{k}}^{(\tilm)}\,,
\end{align}
with $\tilm = -\tilj + 1, ..., \tilj - 1$, 
and the initial condition given by 
$\alpha_{\text{L},\bm{k}}^{(\tilm)}(\eta_i) = 1$ and $\beta_{\text{L},\bm{k}}^{(\tilm)}(\eta_i) = 0$, where
\begin{align}
	\partial_\eta \theta_{\tilm} &= \frac{k}{2f} \frac{\sqrt{\tilj^2 - \tilm^2}}{(k/f + \tilm)^2 + \tilj^2 - \tilm^2}
	\frac{f'}{f}\,,
	\quad
	\Theta_{\tilm} = \int_{\eta_i}^{\eta} d\bar{\eta} 
	\sqrt{(k + \tilm f)^2 + f^2(\tilj^2 - \tilm^2)}\,.
	\label{eq:theta}
\end{align}
In the following we shall work with the Born approximation $\alpha_{\text{L},\bm{k}}^{(\tilm)} \simeq 1$ 
which is valid for $|\beta_{\text{L},\bm{k}}^{(\tilm)}| \ll 1$. 
We will check that this is indeed the case in the end of the computation.\footnote{
	We will see below that $|\beta_{\text{L},\bm{k}}^{(\tilm)}|$ is suppressed at least by $f^2/k^2$
	for the large momentum limit, thus this approximation is enough also for the discussion of the log divergence.
}
Then the Bogolyubov coefficient~$\beta$ is simply given by
\begin{align}
	\beta_{\text{L},\bm{k}}^{(\tilm)}\left(\eta_f\right) 
	\simeq \int_{\eta_i}^{\eta_f}d\eta\, \left(\partial_\eta \theta_{\tilm}\right) e^{-2i \Theta_{\tilm}}\,.
	\label{eq:beta_born}
\end{align}
In the following we solve this equation in the small and large momentum limit, respectively.

\paragraph{Small momentum limit.}
In the limit $k \ll \tilj f_i$, we may approximate Eq.~\eqref{eq:theta} as
\begin{align}
	\partial_\eta \theta_{\tilm} &\simeq \frac{kf'}{2f^2}\sqrt{1 - \frac{\tilm^2}{\tilj^2}}\,, 
	\quad	
	\Theta_{\tilm} \simeq \tilj \int_{\eta_i}^{\eta} d\bar{\eta}\, f\,.
\end{align}
By taking the time evolution of $f(\eta)$ as $f(\eta) \simeq \xi/(-\eta)$,
we easily evaluate the integral~\eqref{eq:beta_born} as
\begin{align}
	\abs{\beta_{\text{L},\bm{k}}^{(\tilm)}}\left(\eta_f\right) \simeq
	\frac{k(-\eta_i)}{2\tilj \xi}\sqrt{\frac{1-\tilm^2/\tilj^2}{1+4\tilj^2\xi^2}}\,.
\end{align}
It satisfies $|\beta_{\text{L},\bm{k}}^{(\tilm)}| \ll 1$, and hence the Born approximation is self-consistent.
By taking $\tilj = 1$ and $\tilm = 0$, it reproduces the upper formula in Eq.~\eqref{eq:beta_approx}.

\paragraph{Large momentum limit.}
In the limit $k \gg \tilj f_f$, we may approximate Eq.~\eqref{eq:theta} as\footnote{
	Here we ignore the next-to-leading order contribution from the combination $(k+\tilm f)$.
	This is because it just shifts the lower bound of the integration,
	and hence is irrelevant for the divergence.
}
\begin{align}
	\partial_\eta \theta_{\tilm} &\simeq \frac{f'}{2k}\sqrt{\tilj^2 - \tilm^2}\,, 
	\quad	
	\Theta_{\tilm} \simeq k\eta\,.
\end{align}
Now it is easy to evaluate Eq.~\eqref{eq:beta_born}.
By taking $f = \xi/(-\eta)$ as usual, we obtain up to the next-to-leading order as
\begin{align}
	\beta_{\text{L},\bm{k}} \simeq 
	\frac{\tilj \xi e^{-2ik\eta}}{4}\sqrt{1 - \frac{\tilm^2}{\tilj^2}}\left(\frac{i}{k^2\eta^2} - \frac{1}{k^3\eta^3} \right)\,,
\end{align}
where we have approximated $k(-\eta_i) \gg 1$.
It again satisfies $|\beta_{\text{L},\bm{k}}^{(\tilm)}| \ll 1$, 
indicating that the Born approximation is self-consistent.
By taking $\tilj = 1$ and $\tilm = 0$, it reproduces the lowest line of Eq.~\eqref{eq:beta_approx} as well as Eq.~\eqref{eq:beta_nlo}.

\small
\bibliographystyle{utphys}
\bibliography{refs_cs-cosmology}{}
  
\end{document}